%% file: SRFReview.tex
\documentclass[aps,reprint,amsmath,amssymbols,superscriptaddress,nofootinbib]{revtex4-1}
\usepackage{epsfig}
\usepackage{refcount}
\usepackage{amsmath}
\usepackage{graphicx}
\usepackage{multirow,hhline}
\usepackage{dcolumn}
\usepackage{natbib}
\usepackage{amssymb}
\usepackage{color, hyperref, upgreek}

\definecolor{darkBlue}{rgb}{0,0,0.6}
\definecolor{darkRed}{rgb}{0.5,0,0}
\definecolor{darkGreen}{rgb}{0,0.5,0}

\newcommand{\Hsh}{\ensuremath{H_\mathrm{sh}}}

\begin{document}

\title{Theoretical estimates of maximum fields in superconducting resonant
radio frequency cavities: Stability theory, disorder, and laminates}

\author{Danilo B. Liarte}
\address{Laboratory of Atomic and Solid State Physics, Clark Hall, Cornell University, Ithaca, New York 14853-2501}

\author{Sam Posen}
\address{Fermi National Accelerator Laboratory, Batavia, IL 60510, USA.}

\author{Mark K. Transtrum}
\address{Department of Physics and Astronomy, Brigham Young University, Provo, Utah 84602, USA}

\author{Gianluigi Catelani}
\address{Forschungszentrum J\"{u}lich, Peter Gr\"{u}nberg Institut (PGI-2), 52425 J\"{u}lich, Germany}

\author{Matthias Liepe}
\address{LEPP, Physics Department, Newman Laboratory, Cornell University}

\author{James P. Sethna}
\address{Laboratory of Atomic and Solid State Physics, Clark Hall, Cornell University, Ithaca, New York 14853-2501}


\begin{abstract}
Theoretical limits to the performance of superconductors in high magnetic fields parallel to their surfaces are of key relevance to current and future accelerating cavities, especially those made of new higher-Tc materials such as Nb$_3$Sn, NbN, and MgB$_2$.
Indeed, beyond the so-called superheating field $\Hsh$, flux will spontaneously penetrate even a perfect superconducting surface and ruin the performance. We present intuitive arguments and simple estimates for $\Hsh$, and combine them with our previous rigorous calculations, which we summarize.
We briefly discuss experimental measurements of the superheating field, comparing to our estimates.
We explore the effects of materials anisotropy and the danger of disorder in nucleating vortex entry.
Will we need to control surface orientation in the layered compound MgB$_2$? Can we estimate theoretically whether dirt and defects make these new materials fundamentally more challenging to optimize than niobium? Finally, we discuss and analyze recent proposals to use thin superconducting layers or laminates to enhance the performance of superconducting cavities.
Flux entering a laminate can lead to so-called pancake vortices; we consider the physics of the dislocation motion and potential re-annihilation or stabilization of these vortices after their entry.
\end{abstract}

\maketitle

%
%
%
%
%

\input{Sec/Intro}

\input{Sec/BasicTheory}

\input{Sec/Experiments}

\input{Sec/FancierTheories}

\input{Sec/Laminates}

\input{Sec/Conclusions}

\section*{Acknowledgment}
Our work on the superheating field was urged upon us by Hasan Padamsee, who recognized both the importance of firm estimates of the theoretical maximum performance for niobium cavities, and the confusion in the SRF field at that time about potential new materials. We also thank Alex Gurevich for extensive consultation and collaboration, particularly on the new work on laminates in Section~(\ref{sec:laminates}). Much of that section was inspired by our conversations with him and tests his independently derived thoughts and conclusions on the subject. DBL and JPS were supported by NSF DMR-1312160, Sam Posen is supported by the United States Department of Energy, Offices of High Energy Physics. Fermilab is operated by Fermi Research Alliance, LLC under Contract No. DE-AC02-07CH11359 with the United States Department of Energy. GC acknowledges partial support by the EU under REA Grant Agreement No. CIG-618258. ML was supported by NSF Award No. PHY-1416318, and DOE Award No. DE-SC0008431.
This work was supported by the U.S. National Science Foundation under Award OIA-1549132, the Center for Bright Beams.


%

\end{document}

%% file: Sec/Intro.tex
\section{Introduction}
\label{sec:intro}

To transfer energy to beams of charged particles, accelerators frequently use superconducting radio-frequency (SRF) cavities, devices that are capable of sustaining large amplitude electromagnetic fields with relatively small input power. The energy gain of a beam traversing a cavity is determined by the electric field amplitude along its path---a larger amplitude can reduce the number of cavities required to reach a given energy. This is especially important in high energy accelerators, which call for as many as tens of thousands of cavities \cite{ILCtdr}. It is therefore of interest to understand the mechanisms that fundamentally limit the accelerating electric field.
For state-of-the-art SRF cavities that have been carefully prepared to prevent non-fundamental degradation processes such as field emission \cite{Bernard1992,Kneisel1993} and multipacting \cite{Proch1979}, studies show that the limit is not the electric field, but rather the interaction of the magnetic field with the superconducting material of the cavity walls.
The fundamental limit to acceleration in SRF cavities is the {\em superheating field}
$\Hsh$, introduced in Section~(\ref{sec:basictheory}).

This article will cover ideas, methods, and results revolving around the superheating field
and its dependence on the superconductor -- materials properties, anisotropy, defects and disorder,
and laminates. The ideas and methods are primarily gleaned from the broader condensed matter 
community. In Section~(\ref{sec:basictheory}) we review computations of $\Hsh$ for clean systems
using field theories from the 1950's derived for pure superconductors near their transition 
temperature~\cite{tinkham96}; in Section~(\ref{subsec:eilenberger}) we draw from more sophisticated
theories from the 1960's to calculate $\Hsh$ at all temperatures~\cite{eilenberger1968}, and discuss
the future need to use these historical theories to incorporate effects of strong 
coupling and electronic structure~\cite{eliashberg60} in new materials. In
Section~(\ref{subsec:anisotropy}) we review the use of these methods to address the electronic
anisotropy of some of the new materials.
In Section~(\ref{subsec:disorder} we introduce an illustrative calculation of 
the effects of disorder using tools and methods developed in the 60's for disordered
systems~\cite{lifshitz64,zittartzL66} and nucleation theory~\cite{langer68,DanielsS11}, providing
reassurance that new materials will likely not be far more sensitive to flaws and dirt. Finally,
in Section~(\ref{sec:laminates} we investigate the properties of superconducting laminates, by
drawing from work from the 90's on the dynamics of `pancake vortices' in certain layered 
high-temperature superconductors~\cite{huebener01} (particularly BSCCO,
Bi$_2$Sr$_2$Ca$_{n-1}$Cu$_n$O$_{2n+4+x}$).

We frankly have two goals for this article. As discussed above, we wish to provide an introduction
for the accelerator community into tools and methods from the broader condensed matter community that
can help interpret current experimental challenges and guide plans for future research in
optimizing materials properties for SRF cavities. But conversely, we want to provide a window
for the broader condensed matter theory community into the remarkable frontiers of field, frequency,
and materials preparation being explored by the SRF community. We invite their participation
in melding 21st century materials-by-design tools from electronic structure theory with
20th century field theories of superconductivity, bridging the scales to address current
technological challenges in the accelerator field. (Full disclosure: this article was supported
in part by the Center for Bright Beams, an NSF Science and Technology Center whose mission
is precisely to bring the accelerator community together with outside experts in
physical chemistry, materials science, condensed matter physics, plasma physics and mathematics.)


\subsection{Basic facts about superconductors: type I and II, $H_c$, $H_{c1}$, and $H_{c2}$}

Normal conducting metals, such as copper, are not viable as radio-frequency cavities for long-pulse high-gradient applications. Due to their high surface resistance, these cavities dissipate too much power on the walls, which can result in melting, among other structural problems, if they are not sufficiently cooled. When subject to high accelerating fields, copper cavities are limited to short-pulse applications. In contrast, superconducting radio-frequency cavities have a much lower surface resistance, which implies low dissipation on the walls and high quality factors (of about $10^{10}$, compared to $10^4$ for copper)~\cite{padamsee09}. Taking into account the refrigerator power to keep the cavity in the superconductor state, SRF cavities are considerably more economical than copper cavities, and present huge benefits, especially for long-pulse applications. At high magnetic fields, however, high-temperature superconductor cavities can dissipate as much power as copper due to the nucleation and motion of vortices. 


At low enough temperature and applied magnetic field (which for now we assume to be constant in time), superconductors exhibit the Meissner effect: magnetic fields are expelled from the interior of the superconductor, exponentially decaying from the interface surface. Larger applied magnetic field can destroy this Meissner state in two ways, depending on the type of superconductor. In type-I superconductors, an abrupt phase transition takes place at the thermodynamic critical field $H_c$, above which the superconductor is in the normal state. In type-II superconductors, the situation is slightly more complicated. Magnetic flux penetration starts, via vortex nucleation, at a lower magnetic field $H_{c1}<H_{c}$. $H_{c1}$ is called the lower critical field. The transition to the normal phase takes place at the upper-critical field $H_{c2}$ ($H_{c2}>H_c$). In the intermediate range, $H_{c1}<H<H_{c2}$, the system is in the vortex lattice state~\footnote{At higher magnetic fields ($>H_{c2}$), \emph{surface} superconductivity can persist up to a third critical field, $H_{c3}$. This critical field should not be mistaken by the \emph{superheating field}, below which the system displays \emph{bulk} superconductivity and field expulsion.}.

\subsection{The superheating field}
\label{sec:superheatingFieldI}

For these cavities during operation, the
external magnetic field is parallel to the superconductor surface. In many applications, the threshold field for flux penetration onto the superconductor is not set by $H_c$ or $H_{c1}$ (for type-I and type-II superconductors, respectively); it is set by the metastability limit of the Meissner state, i.e. by the \emph{superheating} field~\cite{transtrum11,catelani08,garfunkel52,ginzburg58,kramer68,gennes65,galaiko66,kramer73,fink69,christiansen69,chapman95,dolgert96,bean64}. The Meissner state is metastable at $H_c<H<H_{\mathrm{sh}}$ for type-I superconductors, and at $H_{c1}<H<H_{\mathrm{sh}}$ for type-II superconductors. The onset of instability of the Meissner state is related to the vanishing of a surface energy barrier that prevents field penetration onto the superconductor even when $H>H_c$ or $H>H_{c1}$.

The metastable Meissner state is analogous to the state of \emph{superheated} water (perhaps explaining the name ``superheating field''). Liquid water in a glass can be superheated in a microwave to a temperature above the liquid-gas transition temperature, but still remain in the liquid state due to the surface tension barrier at the liquid-gas interface, causing small vapor bubbles to contract rather than grow. Surface tension in water is analogous, for instance, to the surface tension due to the energy barrier preventing vortex nucleation in type-II superconductors. Unlike the case of water, as we argue in Section~(\ref{subsec:thermalFluctuations}), thermal nucleation of vortices occurs at relatively long time scales, suggesting that the Meissner state can be sustained in RF applications for fields as large as the superheating field. However, this scenario can considerably change when one considers the effects of disorder in the superconductor. Section~(\ref{subsec:disorder}) discusses disorder-induced nucleation of vortices.

The superheating field is associated with spinodal curves where the local stability of the Meissner state is broken. This is a more precise definition that is useful for both type-I and type-II superconductors. We shall discuss calculations of the superheating field in Section~(\ref{sec:basictheory}). Our calculations
there will be assuming an external field that is constant in time and ignore
thermal fluctuations. We here discuss these approximations.

\subsection{Why GHz is slow}
\label{sec:slowGHz}

Calculations of the superheating field for DC applied magnetic fields will be accurate for RF applications when the microscopic relaxation times are smaller than the time scales that are associated with changes in the fields inside the cavity. Time scales for the latter are of order of nanoseconds~\cite{padamsee09}. A version of time dependent Ginzburg-Landau theory given by Gor'kov and Eliashberg predicts the characteristic relaxation time near $T_c$: $\tau_{\mathrm{GL}}=\pi \, \hbar / [8 \,k\, (T_c-T)]$, where $\hbar$ is the Planck constant divided by $2\pi$ and $k$ is the Boltzmann constant~\cite{tinkham96}. For $T_c-T=1$K, one obtains $\tau_{\mathrm{GL}} \sim 10^{-3}\,$ns for oscillating fields parallel to the sample surfaces. Using $ \Delta \sim k \, T_c$, where $\Delta$ is the superconductor gap, we find $ \tau_{\mathrm{GL}} \sim \Delta^{-1} $ at low temperatures, which is similar to the scaling of collective modes in unconventional superfluids (see e.g. Section 23.5 of Ref.~\cite{ketterson99}). However, note that Gor'kov and Eliashberg theory is applicable to \emph{gapless} superconductors, filled with magnetic impurities and sufficient pair-breaking strength. For superconductors with a clean gap, the relaxation time is expected to be larger than $\tau_{\mathrm{GL}}$, and to scale with the inelastic phonon-scattering time $\tau_E$, which, near $T_c$ is of the order of $\sim 10^{-8}\,$s in Al and $\sim 10^{-11}\,$s for Pb~\cite{tinkham96}, due to its larger critical temperature~\footnote{A simple estimate given in Sec. 10.3 of Ref.~\cite{tinkham96}, assuming a Debye phonon spectrum and free-electron Fermi surface, gives $\tau_E$ scaling as ${T_c}^{-3}$.}. Yoo et al. measured an ultra fast electron-phonon relaxation time of $360$ fs for niobium~\cite{yoo90}. So,
at GHz frequencies we may ignore the time dependence in studying the
stability.

\subsection{Why thermal fluctuations are small}
\label{subsec:thermalFluctuations}

\renewcommand{\arraystretch}{1.5}
\begin{table*}
\begin{center}
\begin{tabular}{| c | r | r | r | r | r | r | r | r | r | r |}
\hline Material & $\lambda$[nm] & $\xi$[nm]  & $\kappa$ & $T_c$[K] & $H_{c1}$[T] & $H_{c}$[T] & $\Hsh$[T] & $F$[J/m$^3$] & $F\xi^3/k_B$ [K] \\ \hline\hline
Nb &		40 &	27 & 1.5 & 9 &  0.13 &	0.21 &	0.25 &	17547 &	25009.0 \\ \hline
Nb$_3$Sn &	111 &	4.2 & 26.4 & 18 & 0.042 &	0.5 &	0.42 &	99472 &	533.6 \\ \hline
NbN &		375 &	2.9 & 129.3 & 16 & 0.006 &	0.21 &	0.17 &	17547 &	31.0 \\ \hline
MgB$_2$ &	185 &	4.9 & 37.8 & 40 & 0.017 &	0.26 &	0.21 &	26897 &	229.1 \\ \hline
\end{tabular}
\caption{Representative material parameters for niobium, the traditional superconducting material used in SRF cavities, as well as candidate SRF materials that have the potential to reduce cooling costs due to their higher $T_c$. The coherence length $\xi$ is calculated using equations in Ref.~\cite{Orlando1979}. The penetration depth $\lambda$ is calculated from Eq.~3.131 in Ref.~\cite{tinkham96}. The ratio $\kappa=\lambda/\xi$ is called the Ginzburg-Landau parameter, and determines many properties of superconductors. A residual resistivity ratio of 100 was assumed for niobium. For MgB$_2$, the values of $\lambda$ and $\xi$ are experimental values given in the reference. For calculations, $H_c=\phi_0 / [\mu_0(2\sqrt{2}\pi\xi\lambda)]$ is used \cite{tinkham96}. $H_{c1}$ for Nb is found from fit to numerically computed data in Ref.~\cite{Hein1999} and [\onlinecite{Harden1963}]. $H_{c1}$ for strongly type II materials is found from Eq.~5.18 in Ref.~\cite{tinkham96}. $\Hsh$ is calculated using $\Hsh\simeq H_c(0.75+0.54 \, \kappa^{-1/2})$ \cite{transtrum11}. The condensation energy density $F$ is given by $\mu_0H_c^2 / 2$ \cite{tinkham96}. Nb data is extracted from Ref.~\cite{Maxfield1965}, Nb$_3$Sn data from Ref.~\cite{Hein1999}, NbN data from Ref.~\cite{Oates1991}, and MgB$_2$ data from Ref.~\cite{Wang2001}.}
\label{tab:MaterialsParameters}
\end{center}
\end{table*}

One key question for our purposes is whether thermal fluctuations can help
activate vortices over the surface barrier. 
Thermal fluctuations in most superconductors (apart from the high-$T_c$ 
cuprate superconductors) are very small. This is due to the same approximation
that makes the BCS theory of superconductors so successful. BCS
theory is a {\em mean-field theory} of interacting Cooper pairs, which becomes
exact when each Cooper pair interacts with an infinite number of neighbors
(thus seeing the {\em mean} behavior of the system). Each Cooper pair is
of radius roughly the coherence length $\xi$, so BCS theory will be valid
when the density of Cooper pairs times $\xi^3$ is large. Simple estimates show that there are about $10^6$ centers of Cooper pairs within the region occupied by each pair state; a scenario where the pairs strongly overlap in space, and each pair only feels the average occupancy of the other pair states~\cite{schrieffer99}.
Thermal fluctuations of vortices will be unimportant so long as the condensation
energy density---the amount of energy $F$ that is necessary to destroy superconductivity over a unit volume---times $\xi^3$, is large compared to $k_B T$. 
Table~(\ref{tab:MaterialsParameters}) gives the characteristic temperature
$T_{th} = F \xi^3 / k_B$ where fluctuations will become important, 
for niobium and also three candidate materials being
explored for next generation accelerating cavities. Only for NbN is this
characteristic temperature remotely comparable to $T_c$.

We can gain further insight from an analytic calculation of $ E_v / ( k_B \, T ) $, where $E_v$ is the energy per unit length of a vortex line integrated over a coherence length $\xi$. Using results from BCS theory, the zero-temperature thermodynamical critical field is given by $H_c (0) = 2\, \sqrt{\pi} \sqrt{ \mathcal{N}(0)} \Delta$, where $\mathcal{N}(0) = m \, k_F / (2 \, \pi^2 \, \hbar^2)$ is the density of states at the Fermi energy, $\Delta$ is the superconductor gap at zero temperature, and $k_F$ is the Fermi wave number. Also, $\Delta \approx 1.76 \, k_B \, T_c$, and the coherence length $\xi_0 = \hbar v_F / (\pi \Delta)$, where $v_F$ is the Fermi velocity. Thus,
\begin{eqnarray}
\frac{ E_v }{ k_B \, T } \sim \frac{{H_c}^2 \xi^3}{k_B T} \approx  \frac{ 1.4 }{ t } \left( \frac{ \varepsilon_F }{ \Delta } \right)^2,
\end{eqnarray}
where $ \varepsilon_F = \hbar^2 \, {k_F}^2 / ( 2 \, m ) $ is the Fermi energy, and $ t = T / T_c $. Since the gap is much smaller than the Fermi energy, we can neglect thermal nucleation of vortices; unlike the case of superheated water, the effects of thermal fluctuations is very small. More generally, we expect that $\tau_{\mathrm{mic}} \ll \tau_{\mathrm{cav}} \ll \tau_{\mathrm{t.n.v.}}$ within the Meissner metastable state, where $\tau_{\mathrm{mic}}$, $ \tau_{\mathrm{cav}}$, and $\tau_{\mathrm{t.n.v.}}$ correspond to time scales associated with microscopic degrees of freedom, the variation of the cavity fields, and thermal nucleation of vortices, respectively. 

The negligible effects of thermal fluctuations tells us that estimating the
limiting superheating field of a perfectly clean surface will not be analogous
to bubble formation for superheated water. Instead, we shall use {\em linear
stability theory} in Section~(\ref{subsec:linearstability}) to estimate
the field at which the uniform Meissner state becomes energetically unstable
to an infinitesimal perturbation in the space of magnetic fields and 
superconducting order. A variant of critical droplet theory will appear in
Section~(\ref{subsec:disorder}), where we estimate the effects of 
flaws and disorder in nucleating vortex penetration.

%% file: Sec/BasicTheory.tex
\section{Basic theory of the superheating field}
\label{sec:basictheory}


The superheating field $H_{\mathrm{sh}}$ is set by the competition between magnetic pressure (imposed by the external magnetic field), the energy cost to destroy superconductivity, and the attractive force due to the zero-current boundary condition at the interface. In Ginzburg-Landau theory, the ratio $\kappa = \lambda/\xi$ of the penetration depth $\lambda$ to the coherence length $\xi$ determines many properties of superconductors. In particular, $\kappa < 1/\sqrt{2}$ and $\kappa>1/\sqrt{2}$ are associated with type-I and type-II superconductivity, respectively. In the flux-line lattice of type-II superconductors, both the vortex supercurrent and magnetic field are confined to a tube of radius $\lambda$. The superconductivity is destroyed (the density of superconducting electrons vanishes) over a smaller vortex core of radius $\xi$. Within GL theory, $H_{\mathrm{sh}}(T)/H_c(T)$ depends on materials properties only through the parameter $\kappa$, which is independent of temperature. A careful calculation using linear stability analysis~\cite{transtrum11} shows that $H_{\mathrm{sh}}$ plateaus at about $0.75 H_c$ in the large $\kappa$ limit, and diverges as $\kappa^{-1/2}$ for $\kappa \ll 1$.

\subsection{Simple arguments for the superheating field}
\label{subsec:simpleArgument}

We now give simple arguments and pictures to estimate the superheating field of superconductors (see e.g.~\cite{liarte16}). The main idea is to compute the work necessary to push magnetic field onto the superconductor through an energy barrier set by the magnetic energy, and compare the result with the condensation energy. It is worth noting that there are important qualitative differences between these simple arguments and the actual linear stability analysis of the GL free energy. We will return to these issues when we discuss the effects of anisotropy in Sec.~(\ref{subsec:anisotropy}), and discuss them further in the full publication~\cite{liarte16}.

\begin{figure}[!h]
\centering
\includegraphics[width=0.9\linewidth]{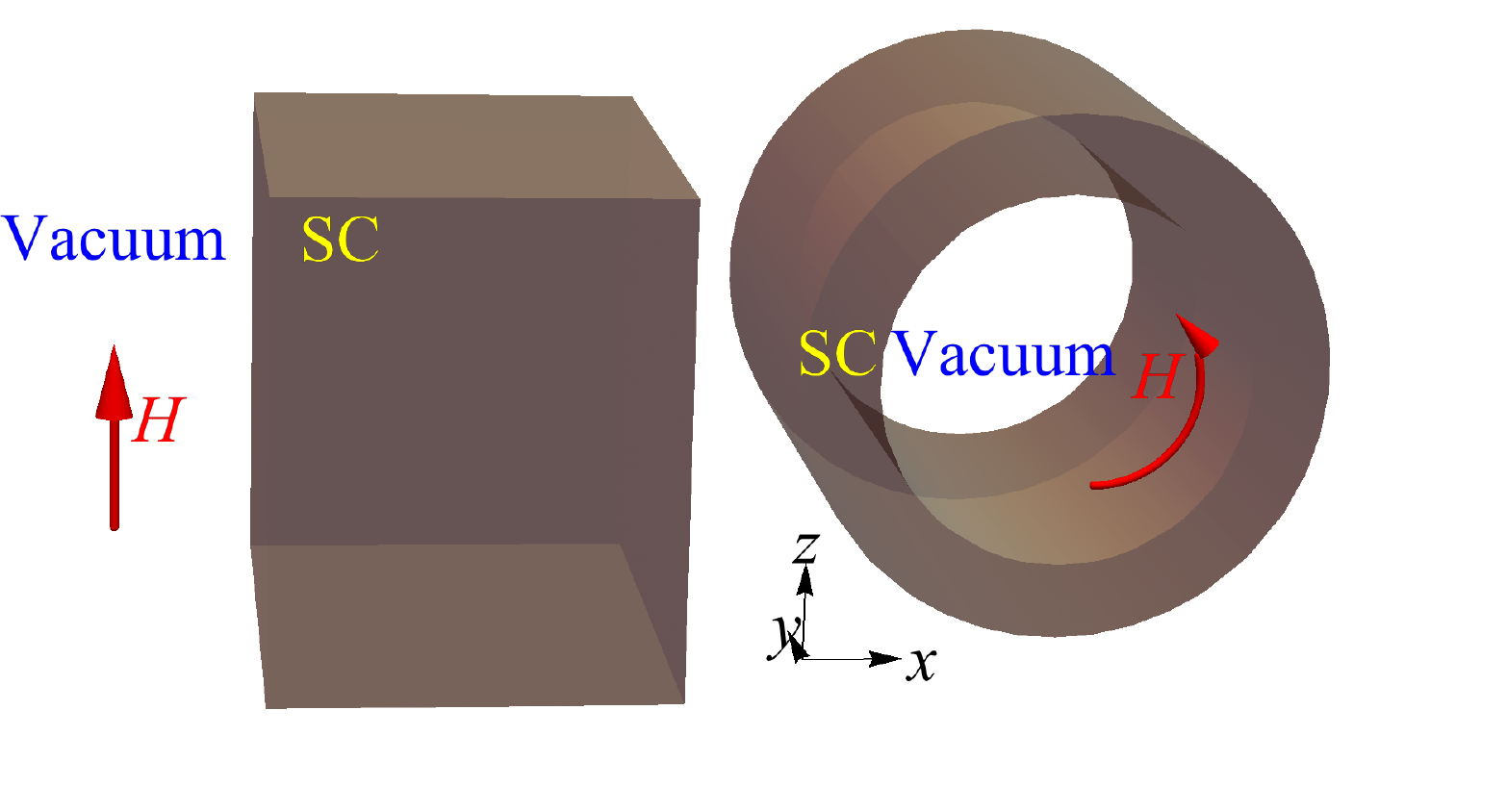}
\caption{(On the left) Illustration of a superconductor occupying the half-space $x>0$, and subject to an applied magnetic field $\mathbf{H}$ that is parallel to the $z$ axis. ``SC'' stands for superconductor. (On the right) Approximate shape of a superconducting RF cavity in the regions of high magnetic fields. As in the flat case, the magnetic field that is generated by the accelerating beam (and excited by an external RF source, driving the operating/accelerating mode) is parallel to the interior surface of the cavity. \label{fig:geometry}}
\end{figure}

Consider a superconductor occupying the half-space $x>0$, and subject to an applied magnetic field $\mathbf{H}$ that is parallel to its surface, along the direction $z$. We illustrate this geometry on the left side of Fig.~(\ref{fig:geometry}), where ``SC'' stands for superconductor. Note that the superconductor region extends to infinite in the positive and negative $y$ and $z$ directions, and in the positive $x$ direction; there are no `corners' in this geometry~\footnote{The absence of corners is an important limiting factor in our
  approach, for corners typically facilitate field penetration in real
  samples of arbitrary shapes. Modern RF cavities have an approximate
  cylindrical shape in the region of high magnetic fields (see right side of 
  Fig.~(\ref{fig:geometry})), with no corners, so such geometric considerations
  become unimportant.}.

Let us start with the argument for the superheating field of a type-I superconductor. For small external magnetic fields, the order parameter does not vanish at the vacuum-superconductor interface. However, if we push a slab of magnetic field onto the superconductor (just enough to make the order parameter vanish at the interface), we will destroy superconductivity over a length scale of order $\xi$. The work per unit area that is necessary to push magnetic energy onto the superconductor is set by the magnetic pressure and the penetration length; it is given approximately by $[H_{\mathrm{sh}} / (4\pi)] H_{\mathrm{sh}} \, \lambda$ in cgs units. To estimate the superheating field, we compare this work with the condensation energy per unit area $[{H_c}^2 / (8\pi)] \, \xi$, resulting:
\begin{eqnarray}
\frac{H_{\mathrm{sh}}}{ H_c} \approx 2^{-1/2} \, \kappa^{-1/2}.
\label{eq:typeI_estimate}
\end{eqnarray}
Equation (\ref{eq:typeI_estimate}) should be compared with the small-$\kappa$ limit of the exact result using Ginzburg-Landau theory~\cite{transtrum11}: $H_{\mathrm{sh}} / H_c \approx 2^{-1/4} \, \kappa^{-1/2}$.

In type-II superconductors, field penetration occurs via vortex nucleation, and the superheating field is set by the magnetic pressure that is necessary to push a vortex through a surface barrier onto the superconductor%
    \footnote{\label{foot:Yogi} Note that this argument is {\em not} related to Yogi's `vortex 
    line nucleation'~\cite{Yogi1977,saito04} estimate of $\Hsh$. The latter, developed to 
    analyze impressive experimental data, was qualitatively incorrect~\cite{transtrum11}. In
    particular, its estimate for the metastable limit $\Hsh$ for large $\kappa$ went {\em below}
    $H_{c1}$, which makes no sense. This
    misled the SRF field for years into ignoring the potential importance of higher $\kappa$ 
    materials.}.
There are two steps to this penetration. First, the core of the 
superconducting vortex (of radius $\sim\xi$) must penetrate into the 
surface, at a cost given by the core volume times the condensation energy.
Second, this newly penetrated vortex must fight past an attractive force
toward the surface due to the boundary conditions at the surface,
 which is usually estimated~\cite{bean64} by the attraction
to an `image vortex'. Below we discuss the superheating field 
estimated from the initial penetration of the vortex. (Bean and Livingston's
original estimate~\cite{bean64} of the superheating field starts
(somewhat arbitrarily) at a distance $x=\xi$ after this initial penetration,
and focuses on the effects of the attractive longer-range force.)

\begin{figure}[!h]
\centering
\includegraphics[width=0.9\linewidth]{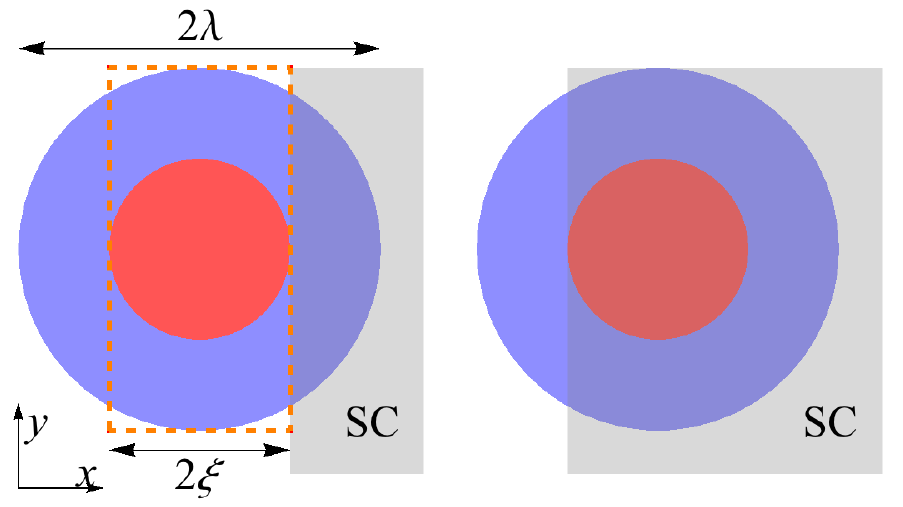}
\caption{Illustrating the penetration of a vortex core into a type-II superconductor (from~\cite{liarte16}). We estimate the superheating field from the work necessary to push a vortex core a distance $x\sim\xi$ into the superconductor. 
The vortex then must fight past an attractive force to a depth
$x\sim\lambda$ to destroy the Meissner state.
\label{fig:vorticesXY}}
\end{figure}

Figure~(\ref{fig:vorticesXY}) illustrates the penetration of a vortex core (red disk) onto a superconductor occupying the half-space $x>0$. The magnetic work per unit length to push the vortex core onto the superconductor is given approximately by the condensation energy (per unit length):
\begin{eqnarray}
\frac{ H_{\mathrm{sh}} }{4\, \pi} \frac{\Phi_0}{\pi \lambda^2} \, 4 \, \lambda \, \xi \approx \frac{{H_c}^2}{8\,\pi} \pi \xi^2,
\label{eq:typeII_estimate_a}
\end{eqnarray}
where $ H_{\mathrm{sh}} / ( 4\, \pi )$ is the magnetic pressure, $\Phi_0$ is the fluxoid quantum, $\pi \lambda^2$ is the vortex area in the $xy$ plane, $4 \, \lambda \, \xi$ is approximately the area that is associated with the region of field penetration (area of the orange box in Fig.~(\ref{fig:vorticesXY}); it is the amount of the area of the vortex that penetrates the superconductor when a vortex core is pushed inside), and $\pi \xi^2$ is the area of the vortex core. Using $\Phi_0 = 2 \, \sqrt{2} \, \pi H_c \, \lambda \, \xi$ in Eq.~(\ref{eq:typeII_estimate_a}):
\begin{eqnarray}
\frac{H_{\mathrm{sh}}}{ H_c} \approx \frac{ \sqrt{2} \, \pi }{32} \approx 0.14,
\label{eq:typeII_estimate_b}
\end{eqnarray}
independent of $\kappa$.

How does this estimate compare with the field estimated from the attractive
force, and with the true answer? The true answer, given below in
Section~(\ref{subsec:linearstability}), is about 
five times larger: $H_{\mathrm{sh}} / H_c \approx 0.75$. Bean 
and Livingston's estimate of the superheating field due to the attractive
force to the image vortex is $H_{\mathrm{sh}}/ H_c = 0.71$, of the
same form as our estimate $0.14$ but larger and closer to the true estimate.
We present the calculation of the field necessary to introduce the core
primarily due to its simplicity, and also because it motivates our 
analysis of anisotropic superconductors in Section~(\ref{subsec:anisotropy}).

One should think of these two contributions as being sequential rather than
serial: first the core must penetrate, and then the vortex must fight
the longer-range attraction to enter the bulk. (It is interesting and convenient
that these two fields are of the same scale.) The GL calculation
in Section~(\ref{subsec:linearstability}) of course incorporates both the
initial core penetration and the longer range attractive force, together 
with cooperative effects of multiple vortices entering at the same time.

Note that, while the {\em field} needed to push the vortex core into
the superconductor is roughly comparable to that needed to push the
vortex past the attractive long-range potential, the two contributions
contribute very differently to the total {\em energy barrier} to flux
penetration. Energy is force times distance: the two forces are comparable
but the Bean-Livingston force acts on a scale longer by a factor
$\kappa = \lambda/\xi$ than our core nucleation, and will dominate
the barrier height.
Finally, note that in practice the dominant mechanisms for vortex nucleation
that set the superheating field will not involve straight vortices penetrating
all along their lengths (as in our calculation above) or, even more
impressively, arrays of straight vortices cooperatively pushing their
way through the surface barrier (Section~(\ref{subsec:linearstability}) below).
We expect that disorder and flaws (discussed in Section~(\ref{subsec:disorder}))
will lead to localized intrusions of single vortex loops into the material
(Fig.~(\ref{fig:fluxTube})).

\subsection{Linear stability calculation of the superheating field}
\label{subsec:linearstability}

In this section we have seen that the superheating field arises in a
bulk superconductor due to the competing effects of magnetic pressure
and the destruction of superconductivity.  Using relatively simple
arguments, we derived the qualitative dependence of this field on $\kappa$.  
We now describe a more rigorous calculation of the superheating field
using a linear stability analysis.  Linear stability analysis is commonly
used in a variety of pattern formation
problems\cite{langer1980instabilities, brower1983geometrical,
  sharma1998pattern, sivashinsky1983instabilities,
  bodenschatz2000recent, ben1984boundary}. For type II
superconductors, the transition from the Meissner state to the mixed
state is triggered by fluctuations of a critical wavelength that
spontaneously break the transverse symmetry of the bulk sample, which
when coupled to the inhomogeneous depth dependence of the Meissner
state, make the superheating transition a challenging application of
this method.  We here describe this calculation using the
Ginzburg-Landau theory for concreteness, although the basic procedure
could be extended to other theories as we discuss below.  Our
presentation follows closely the procedure described in
\cite{transtrum11}, however, the calculation has a long
history in the literature\cite{gennes65, galaiko66,
  kramer68, fink69, christiansen69,
  kramer73, chapman95,
  dolgert96}.

The Ginzburg-Landau free energy for a superconductor occupying the
half space $x>0$ in terms of the magnitude of the superconducting
order parameter $f$ and the gauge-invariant vector potential
$\mathbf{q}$ is given by
\begin{align}
\mathcal{F}[f,\mathbf{q}] & =  \int_{x>0}d^{3}r\Big\{\xi^{2}(\nabla f)^{2}+\frac{1}{2}(1-f^{2})^{2}
 +f^{2}\mathbf{q}^{2}
 \nonumber \\ & \quad
 +(\mathbf{H}_{a}-\lambda\nabla\times\mathbf{q})^{2}\Big\},
\end{align}
where $\mathbf{H}_{a}$ is the applied magnetic field (in units of $\sqrt{2}H_{c}$).

We take the applied field to be oriented along the $z$-axis
$\mathbf{H}_{a} = (0,0,H_a)$, and the order parameter $f=f(x)$ to
depend only on the distance from the superconductors surface.
Assuming that the order parameter is real and parameterizing the vector
potential as $\mathbf{q}=(0,q(x),0)$ fixes the gauge. The
Ginzburg-Landau equations that extremize $\mathcal{F}$ with respect to
$f$ and $\mathbf{q}$ are
\begin{equation}
\xi^{2}f''-q^{2}f+f-f^{3}=0,\qquad
\lambda^{2}q''-f^{2}q=0,
\label{eq:1DGL}
\end{equation}
and with our choices $ H=\lambda q' $, where primes denote derivatives
with respect to $x$.  With appropriate boundary conditions\cite{tinkham96,transtrum11}
these equations can be solved numerically to characterize the Meissner
state.

For a given solution $(f,\mathbf{q})$ we next consider the second variation
of $\mathcal{F}$ associated with small perturbations
$f\rightarrow f+\delta f$ and
$\mathbf{q}\rightarrow\mathbf{q}+\delta\mathbf{q}$ given by
\begin{eqnarray}
\delta^{2}\mathcal{F} & = & \int_{x>0}d^{3}r\Big\{\xi^{2}(\nabla\delta f)^{2}+
4f\delta f\mathbf{q}\cdot\delta\mathbf{q}+f^{2}\delta\mathbf{q}^{2}\nonumber \\
&  & (3f^{2}+\mathbf{q}^{2}-1)\delta f^{2}+\lambda^{2}(\nabla\times\delta\mathbf{q})^{2}\Big\}.
\label{eq:d2F}
\end{eqnarray}
If the expression in Eq.~(\ref{eq:d2F}) is positive for all possible
perturbations, then the solution is (meta) stable. Since the solution
$(f,\delta\mathbf{q})$ depends only on the distance from the boundary
(and is therefore translationally invariant along the $y$ and $z$
directions), we can expand the perturbation in Fourier modes parallel
to the surface. As shown in Ref.~\cite{kramer68}, we can restrict
our attention to perturbations independent of $z$ and write
\begin{align}
&  \delta f (x,y) =\delta\tilde{f}(x)\cos ky, 
\nonumber \\ &
  \delta\mathbf{q}(x,y) =(\delta\tilde{q}_{x}\sin   ky,\delta\tilde{q}_{y}\cos ky,0), \label{eq:fouriermodes}
\end{align}
where $k$ is the wave-number of the Fourier mode. The remaining
Fourier components (corresponding to replacing $\cos \to \sin$ and
vice-versa in Eq.~(\ref{eq:fouriermodes})) are redundant as they
decouple from those given in Eq.~(\ref{eq:fouriermodes}) and satisfy the
same differential equations derived below.

After substituting into the expression (\ref{eq:d2F}) for the second
variation and integrating by parts, we arrive at
\begin{widetext}
\begin{align}
&\delta^{2}\mathcal{F}=\int_{0}^{\infty}dx\left(\begin{array}{ccc}
 \delta\tilde{f} & \delta\tilde{q}_{y} & \delta\tilde{q}_{x}\end{array}\right)
\left(\begin{array}{ccc}
-\xi^{2}\frac{d^{2}}{dx^{2}}+q^{2}+3f^{2}+\xi^{2}k^{2}-1 & 2fq & 0\\
2fq & -\lambda^{2}\frac{d^{2}}{dx^{2}}+f^{2} & -\lambda^{2}k\frac{d}{dx}\\
0 & \lambda^{2}k\frac{d}{dx} & f^{2}+\lambda^{2}k^{2}\end{array}\right)\left(\begin{array}{c}
\delta\tilde{f}\\
\delta\tilde{q}_{y}\\
\delta\tilde{q}_{x}\end{array}\right).\label{eq:d2FOperator}
\end{align}
\end{widetext}
This matrix operator is self-adjoint, and the second variation will be
positive definite if its eigenvalues are all positive.  In the
eigenvalue equations for this operator, the function
$\delta\tilde{q}_{x}$ can be solved for algebraically. The resulting
differential equations for $\delta\tilde{f}$ and $\delta\tilde{q}_{y}$
are
\begin{eqnarray}
-\xi^{2}\delta\tilde{f}''+(3f^{2}+q^{2}-1+\xi^{2}k^{2})\delta\tilde{f}+2fq\delta\tilde{q}_{y}
&=& E\delta\tilde{f}, \nonumber \\ & &
\label{eq:eigen1}\end{eqnarray}
and
\begin{equation}
-\lambda^{2}\frac{d}{dx}\left[\frac{f^{2}-E}{f^{2}+\lambda^{2}k^{2}-E}\delta\tilde{q}_{y}'\right]
+f^{2}\delta\tilde{q}_{y}+2fq\delta\tilde{f}  =  E\delta\tilde{q}_{y},
\label{eq:eigen2}
\end{equation}
where $E$ is the stability eigenvalue. Note that by decomposing in
Fourier modes, the two-dimensional problem is transformed into a
one-dimensional eigenvalue problem. Numerically, it can be solved by
the same methods as the Ginzburg-Landau
equations\cite{transtrum11}.

\begin{figure}
(a) \par\smallskip
\centering
\includegraphics[width=0.9\linewidth]{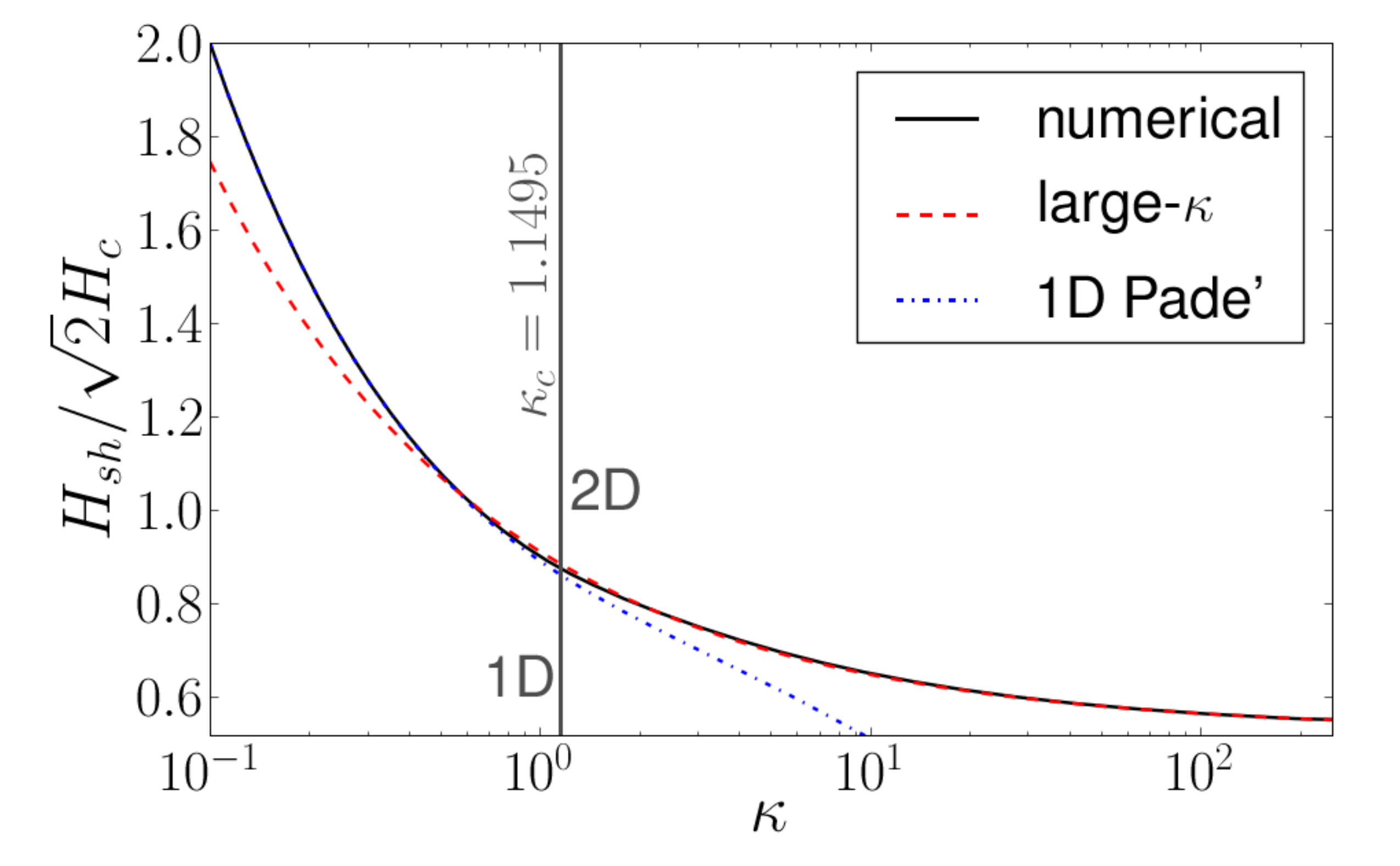}
\\
(b) \par\smallskip
\centering
\includegraphics[width=0.9\linewidth]{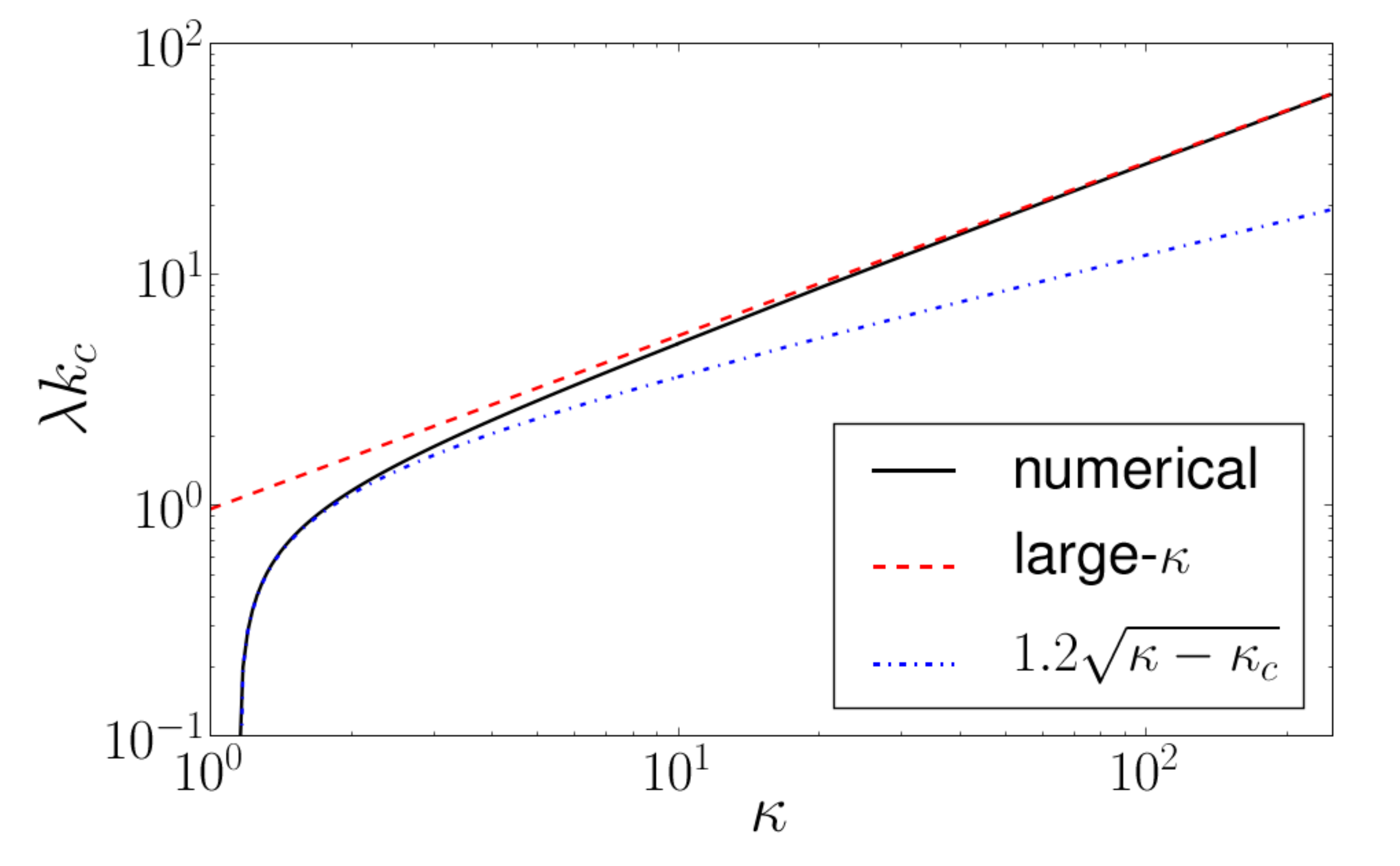}
  \caption{\label{fig:HshGLT} \textbf{Superheating Field in
      Ginzburg-Landau Theory} (from~\cite{transtrum11}). \textbf{(a)} A numerical estimate of
    $\Hsh$  in Ginzburg-Landau theory over many orders of magnitude of
    $\kappa$ was found in reference\cite{transtrum11}
    (black solid line), along with a large-$\kappa$ expansion (red
    dashed line).  A Pad{\'{e}} approximation for small $\kappa$ was derived
    in reference\cite{dolgert96}(blue dotted-dashed line).
    \textbf{(b)} The linear stability calculation also yields the
    wavenumber of the destabilizing fluctuation $k_c$ (black solid line).  This first
    becomes nonzero at $\kappa_c\approx 1.1495$ where it empirically
    behaves like $1.2\sqrt{\kappa - \kappa_c}$ (blue dotted-dashed line).  Large-$\kappa$
    estimates for $k_c$ were also derived in
    reference~\cite{transtrum11} (red dashed line).}
\end{figure}

The stability eigenvalue will depend on the solution of the
Ginzburg-Landau equations, i.e., the applied magnetic field $H_{a}$,
and the Fourier mode $k$ under consideration. The superheating field
is found by varying both the applied magnetic field and Fourier
mode until the smallest eigenvalue first becomes negative.  The
wave-number of the destabilizing fluctuations are therefore found
simultaneously with $\Hsh$  and denoted by $k_c$.  Values of $\Hsh$  and
$k_c$ were calculated in Ginzburg-Landau theory for a wide range of
$\kappa$ in references\cite{transtrum11,
  dolgert96} along with analytic estimates.  The results
are summarized in Figure~(\ref{fig:HshGLT}).

For small $\kappa$, the critical fluctuation occurs with wavenumber
$k_c =0$ while for large $\kappa$, $k_c > 0$.  Interestingly, the
transition to nonzero $k_c$ occurs at some critical $\kappa_c$ that is
distinct from the type-I/type-II boundary ($\kappa = 1/\sqrt{2}$).
Estimates of $\kappa_c$ vary in the literature from
0.5\cite{kramer68} to
1.13($\pm 0.05$)\cite{christiansen69}.  Estimates of
$\kappa_c$ from solving Eqs.~(\ref{eq:eigen1})and (\ref{eq:eigen2})
range from 1.10\cite{fink69} to
1.1495\cite{transtrum11} (our high-accuracy result).

The linear stability approach described in this section could be
extended to other geometries as was done for the case of a
superconducting film separated from a bulk superconductor by a thin
insulating layer in reference~\cite{posen15}.  
More complicated theories of superconductivity can also be solved using
our methods by
replacing the Ginzburg-Landau free energy with the appropriate analog,
such as the Eilenberger formalism described in 
more detail in Section~(\ref{subsec:eilenberger}).

%
%
%
%
%
%

%% file: Sec/Experiments.tex
\section{Experiments}
\label{sec:expt}

\subsection{High Power pulsed RF experiments}
\begin{figure}[!h]
\centering
\includegraphics[width=\linewidth]{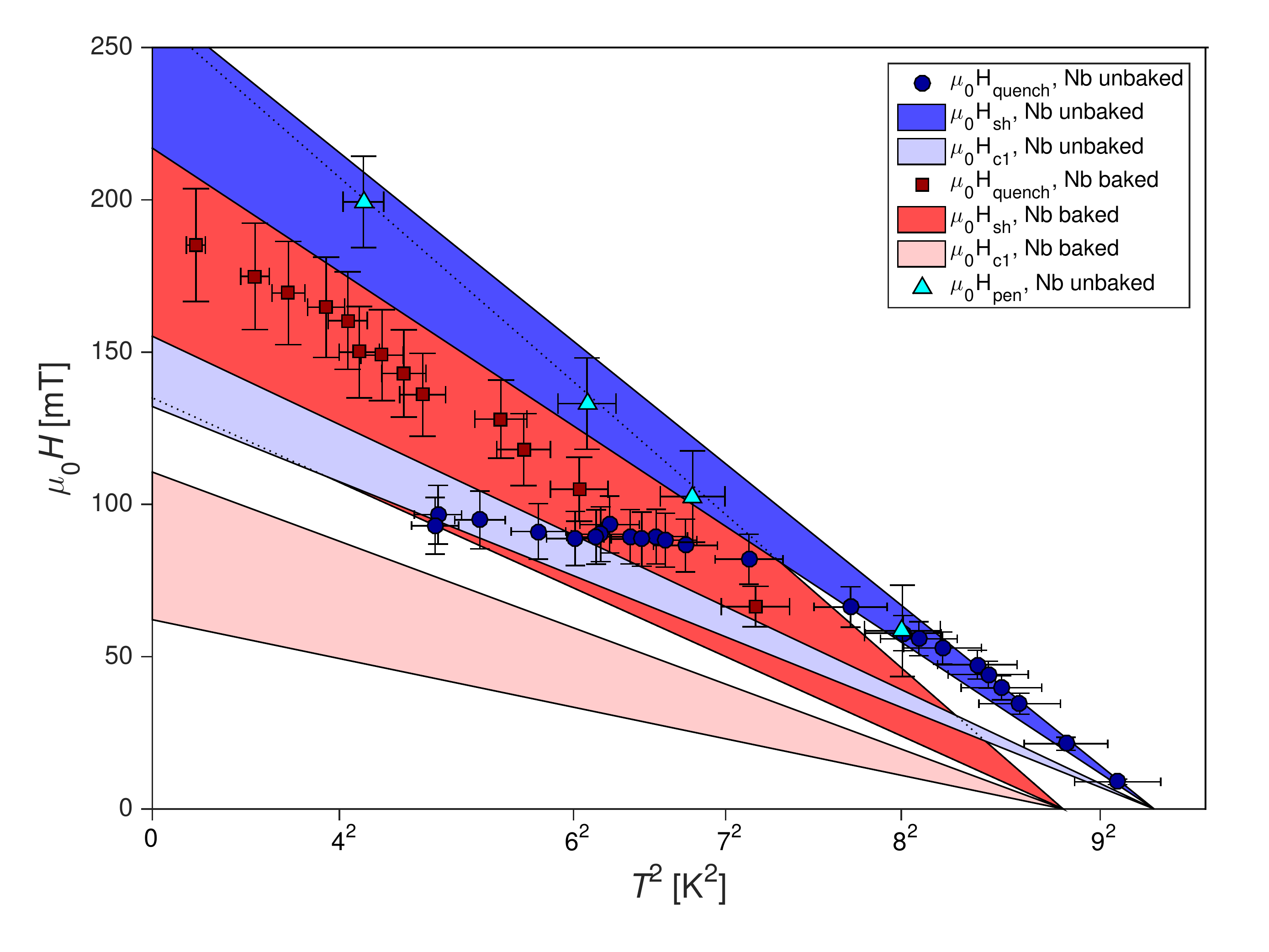}
\caption{Pulsed measurement of the maximum field of superconducting niobium cavities from Valles 
(symbols), compared with estimates of the theoretical maximum possible superheating field (colored ranges). All measurements show good agreement with $\Hsh$ at high temperatures. The cavity baked to removed HFQS degradation (red squares) also shows good agreement at low temperatures. DC flux penetration measurements (green triangles) show good agreement with $\Hsh$ as well. \label{fig:pulsed}}
\end{figure}

Some of the earliest measurements showing $\Hsh>H_c$ for niobium were reported by Renard and Rocher based on DC magnetization measurements. Yogi et al. performed a more systematic study at RF frequencies on samples of Sn, In, Pb, and alloys, in order to cover a range of $\kappa$ values~\cite{Yogi1977}.
Analysis of their data resulted in the vortex line nucleation model discussed in footnote~\ref{foot:Yogi}.
Noting that measurements of the RF critical field have shown inconsistency, Campisi used a very high power RF source at SLAC to very quickly ramp up the fields in cavities \cite{Campisi1985}. The goal of these high power RF measurements is to reduce the influence of defects by outpacing the thermal effects they cause. Campisi performed high power RF measurements on Nb, Nb$_3$Sn, and Pb cavities. Hays and Padamsee performed similar measurements on these materials at Cornell \cite{Hays1997}. The niobium results are reproduced in Fig.(\ref{fig:pulsed}), showing fairly reasonable agreement with the expected superheating field close to $T_c$ \footnote{$T_c$ assumed to be 9.2 K for Valles' data.}, but then diverging at lower temperatures.

After these experiments were performed, new preparation techniques were developed for niobium cavities, including a recipe involving electropolishing and a bake at 120 C. This recipe was found to avoid the ``high field $Q$-slope'' (HFQS) degradation mechanism that occurs in niobium cavities at peak fields of approximately 100 mT \cite{Visentin1998,Kneisel1999}. Experiments by Valles show that pulsed measurements of unbaked niobium produced curves that diverged from the expected $\Hsh$ near the expected onset field of HFQS. However, after the bake was performed, the data agreed very well \cite{Posen2015b}. The $H_{c1}$ and $\Hsh$ curves plotted in the figure were calculated from niobium material parameters that were extracted from measurements of $R_s$ vs $T$ and $f$ vs $T$ via the SRIMP Matthis-Bardeen code \cite{Halbritter1970a,Halbritter1970}. The baked curve has a lower $\Hsh$ due to the change in the mean free path after the bake, which in turn affects $\kappa$.

\subsection{DC flux penetration measurements by N. Valles.}
Valles also performed measurements of the superheating field of unbaked niobium using a DC probe to avoid the effects of HFQS. Using a superconducting solenoid, he applied a DC field to the exterior of a niobium cavity operating at low fields. A sudden decrease in the quality factor of the cavity indicated that flux from the magnet had penetrated to the interior cavity surface. The penetration field extracted from measurements of the applied field agreed well with the expected superheating field for unbaked niobium, as shown in Fig.~(\ref{fig:pulsed}) \cite{Posen2015b}.

%% file: Sec/FancierTheories.tex
\section{Beyond Ginzburg-Landau: Eilenberger, anisotropy, and disorder}
\label{sec:fancier}

The isotropic Ginzburg-Landau analysis of Section~(\ref{sec:basictheory})
is a trustworthy estimate for the superheating field only for ideal surfaces
of single-band superconductors with cubic symmetry near the superconducting
transition temperature $T_c$. In this section we pursue three topics that
introduce new physics to this calculation. First, superconducting RF
cavities are usually run at
temperatures significantly lower than $T_c$; niobium cavities, with 
$T_c \sim 9$K are usually run at $T=2$--$4$K in working accelerators. In 
Section~(\ref{subsec:eilenberger}) we review calculations of the
superheating fields that use Eilenberger theory, which is valid at lower
temperatures,
presenting the analytic results~\cite{catelani08}
at large $\kappa$. 
Our estimates suggest that
these Eilenberger corrections to GL are quantitatively important at
operating temperatures, but not large. Second,
many of the potential
new superconductors have rather anisotropic crystal structures and electronic
properties; if the superheating field has significant anisotropy, this could
motivate single-crystal or controlled growth conditions to control surface
orientations in cavities.
In Section~(\ref{subsec:anisotropy}) we review
calculations~\cite{liarte16} which show that this anisotropy will be small
near $T_c$; we also discuss conflicting results for the anisotropy of
multi-band superconductors (like MgB$_2$) at low temperatures.
Third, in
Section~(\ref{subsec:disorder}) we estimate the effects of disorder and 
flaws in these materials, presenting both qualitative and simple quantitative
estimates of the effects of defects and dirt in locally lowering the
barrier to magnetic flux penetration and thus lowering the effective
superheating field.

\subsection{Eilenberger theory for lower temperatures}
\label{subsec:eilenberger}

The Ginzburg-Landau approach to superconductivity is generally accurate near the critical temperature $T_c$, but usually the accuracy of its prediction worsen as temperature is lowered below $T_c$. A
basic example of its failure is given by the temperature dependence of the order parameter $\Delta$: according to GL theory, $\Delta(T)$ behaves as $\sqrt{1-T/T_c}$, in
agreement near $T_c$ with the microscopic BCS theory. The latter, however, predicts that at low temperatures the order parameter is temperature-independent up to exponentially
small corrections. For our purposes, the limited validity of GL theory implies that the dependence of the superheating field on $\kappa$ discussed in Sec.~(\ref{sec:basictheory}) cannot be assumed to be quantitatively accurate at the low temperatures at which RF cavities are usually operated. This motivates us to consider a more general approach, valid at arbitrary temperature.

For low-$T_c$ superconductors, the coherence length $\xi_0 = \hbar v_F/2\Delta_0$ is much longer than the Fermi wavelength; here $\xi_0$ is the zero-temperature coherence length
for a clean superconductor with zero-temperature order parameter $\Delta_0$ and Fermi velocity $v_F$. Thanks to the separation in length scales (or equivalently, the separation in
energy scales between $\Delta_0$ and the much larger Fermi energy), these
superconductors can be modeled using the so-called \textit{quasiclassical} approach, reviewed for example in Refs.~\cite{chandrasekhar2008,rammer2007}. This powerful approach is quite flexible,
permitting in principle to include effects such as Fermi surface anisotropy and impurity scattering (we will comment on the latter at the end of this section).
This come at the price of having to calculate various Green's functions
from which physical quantities such as the order parameter and the current can be obtained. Such calculations are usually much more involved that those of the GL approach.

It was shown by Eilenberger~\cite{eilenberger1968} that one can arrive at an expression for the thermodynamic potential as functional of order parameter $\Delta(\boldsymbol{r})$ and vector potential $\boldsymbol{A}(\boldsymbol{r})$, similar to the GL functional, once the quasiclassical equations for the Green's function have been solved. While a general
solution is not possible, for the case of a clean superconductor with spherical Fermi surface  we developed in Ref.~\cite{catelani08} a perturbative approach valid for large $\kappa$. Then the thermodynamic potential $\Omega$ is
\begin{align}
\Omega &= \nu \int\!d^3r \left\{\frac{1}{3}\left(\boldsymbol{\nabla}\times\boldsymbol{A}-\boldsymbol{H}_a\right)^2 + \Delta^2 \log \left(\frac{T}{T_\mathrm{c}}\right) \right.
  \nonumber \\ & \quad
  + \!\int\!(dn)\!\left.\left[\frac{\Delta^2}{\omega_n} - 2 \left(\sqrt{\Omega^2_n +\Delta^2} -\omega_n \right) 
  \right. \right. \nonumber \\ & \quad \left. \left.
  +\frac{1}{\kappa_0^2}\frac{\sqrt{\Omega^2_n+\Delta^2}}{4\Omega^2_n}
 \left(\boldsymbol{n}\cdot\boldsymbol{\nabla} s^{(0)}\right)^2\right]\right\} . \label{tp}
\end{align}
In this expression $\nu$ is the density of states at the Fermi energy, lengths are in units of the zero-temperature penetration depth $\lambda_0$,
\begin{equation}
\frac{1}{\lambda_0}  =\frac{8\pi}{3} \left(\frac{2\pi\xi_0}{\Phi_0}\right)^2 \nu \Delta_0^2 \, ,
\end{equation}
the vector potential is in units of $\Phi_0/2\pi\xi_0$ with $\Phi_0$ the magnetic flux quantum, $\kappa_0 = \lambda_0/\xi_0$, and $\boldsymbol{n}$ is the unit vector on
the Fermi surface. We also use the short-hand notations
\begin{equation}
\int\!(dn) = 2\pi T \sum_n \int \frac{d\boldsymbol{n}}{4\pi}\, , \qquad \Omega_n = \omega_n - i \boldsymbol{n}\cdot\boldsymbol{A} \, ,
\end{equation}
with $\omega_n = 2\pi T(n+1/2)$, $n=0,\,1,\,2,\ldots$, the fermionic Matsubara frequencies, and
\begin{equation}
s^{(0)} = \frac{2\Delta}{\sqrt{\Omega_n^2 + \Delta^2}} \, .
\end{equation}

\begin{figure}
\begin{center}
\includegraphics[width=0.9\linewidth]{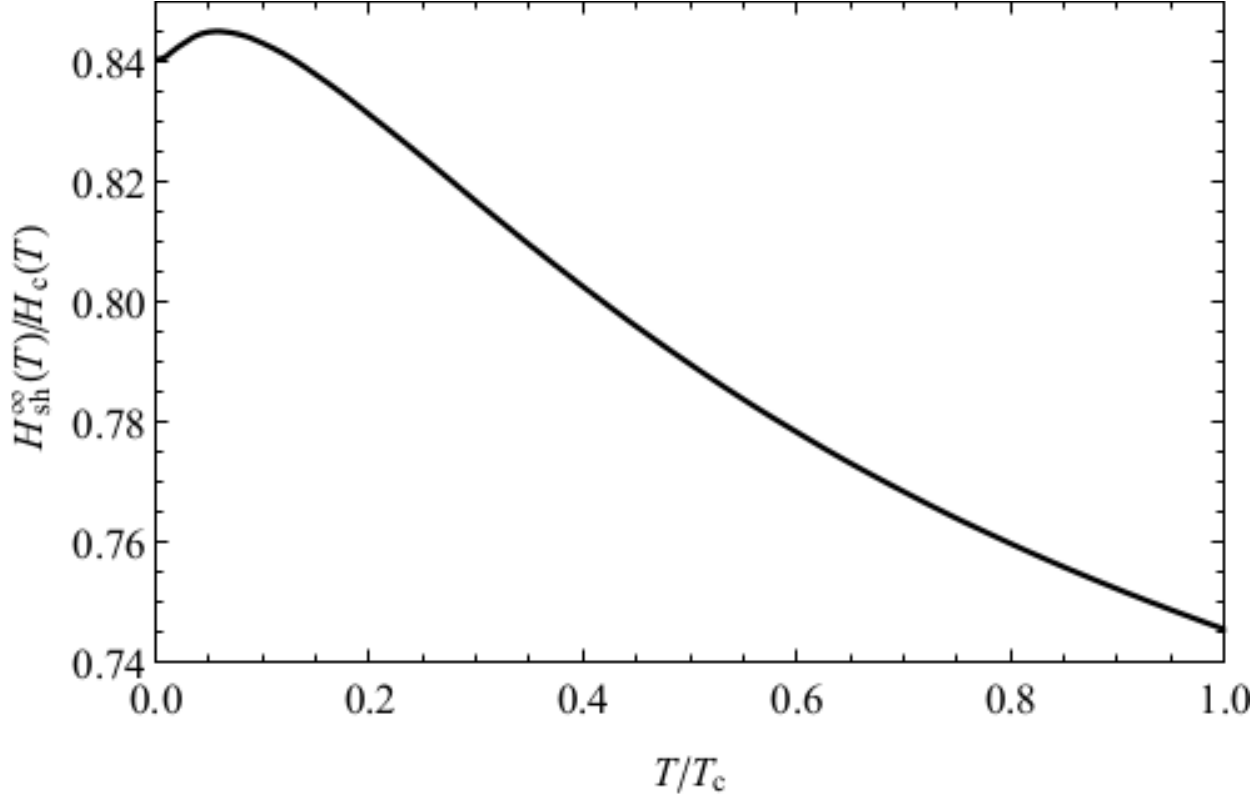}
\end{center}
\caption{Temperature dependence of the ratio $H_\mathrm{sh}^\infty/H_\mathrm{c}$ (from~\cite{catelani08}). Note the non-monotonic behavior at low temperatures.}
\label{fig:eilen}
\end{figure}

The thermodynamic potential in Eq.~(\ref{tp}) reduces to the GL one near $T_c$%
   \footnote{In considering the limit $T\to T_c$ in Ref.~\cite{catelani08},
a prefactor was missed in Eq.~(29) and consequently Eq.~(31), which should read respectively: $\kappa_{GL} = 2\pi T\sqrt{2/3\zeta}\, \kappa_0 \approx 1.50 \kappa_0$
and $\xi(T)=\sqrt{2/3} [\Delta_0/\Delta(T)] \xi_0$ in the notation of that work.} and it can be used to find the superheating field at arbitrary temperature in the regime
$\kappa_0 \gg 1$. The calculation of $H_\mathrm{sh}$ proceeds in the same manner as in the GL approach, by studying the stability against small perturbation
of the local minima of $\Omega$. This study was performed in Ref.~\cite{catelani08} at leading order in $\kappa_0 \to +\infty$. The ratio $H_\mathrm{sh}/H_c$
between superheating and critical field can be calculated analytically at $T=T_c$ and $T=0$:
\begin{equation}
\frac{H_\mathrm{sh}^\infty}{H_c}(T_c) \simeq 0.745 \, , \qquad \frac{H_\mathrm{sh}^\infty}{H_c}(0) \simeq 0.840
\end{equation}
where we use the $\infty$ symbol in the superscript to indicate that these are leading-order results. Interestingly, the zero-temperature ratio is almost 13~\% larger than the near-$T_c$ one, indicating that naive extrapolation to low-temperatures of the GL result underestimates the superheating field. 
At arbitrary temperature, the $H_\mathrm{sh}^\infty/H_c$ ratio can be found numerically and is shown in Fig.~(\ref{fig:eilen}). Note the non-monotonic dependence of $H_\mathrm{sh}^\infty/H_c$ on temperature, which leads the superheating field to acquire its largest value $H_\mathrm{sh}^\infty \simeq 0.843 H_c(0)$ at $T\simeq 0.04 T_c$.

It should be noted that while the Meissner state remains metastable up to $H_\mathrm{sh}$, a clean superconductor can become \textit{gapless} at
a lower field $H_g$~\cite{lin2012effect}; for example at $T=0$ we have $H_g \simeq 0.816 H_c < H_\mathrm{sh}$. The field $H_g$ is relevant to
applications such as superconducting cavities because as the applied field approaches $H_g$, AC losses rapidly increase. Indeed, in the presence
of a gap the AC losses are in general exponentially suppressed, but this ``protection'' from losses is absent in the gapless state.

The above results are restricted to the leading order in $1/\kappa_0$, which makes it possible to neglect the contributions from the last
term in square brackets in Eq.~(\ref{tp}). At next to leading order, that term must be taken into account and leads to an expression for the superheating
field of the form~\footnote{This formula can be obtained by extending to next-to-leading order the calculations of Ref.~\cite{catelani08}
(G. Catelani, unpublished).}
\begin{equation}
\frac{H_\mathrm{sh}}{H_c}(T) \simeq \frac{H_\mathrm{sh}^\infty}{H_c} (T) + \frac{h(T)}{\sqrt{\kappa_0}}.
\end{equation}
This formula, with a weakly temperature-dependent dimensionless coefficient $h(T)$, has the same inverse square root dependence on
$\kappa_0$ as the GL expression~\cite{transtrum11}.

In closing this section, let us comment briefly on the effect of impurity scattering. Both non-magnetic and magnetic impurities were
considered in Ref.~\cite{lin2012effect} in the limit $\kappa\to\infty$. At sufficient strength of the non-magnetic impurities scattering rate,
there are some qualitative changes: the non-monotonicity of $H_\mathrm{sh}(T)$ is suppressed, and more importantly the gap remains open up
to $H_\mathrm{sh}$. However, quantitatively the value of $H_\mathrm{sh}/H_c$ is changed by at most a few percent. In contrast, adding magnetic
impurities strongly decreases $H_\mathrm{sh}$, similar to the well-known suppression of $T_c$ due to the pair-breaking effect of such impurities.

\subsection{Anisotropic superconductors}
\label{subsec:anisotropy}

Layered superconductors can display highly anisotropic critical fields. The upper-critical field of BSCCO,%
  \footnote{The cuprate superconductors have d-wave order parameters, and hence   have an anisotropic gap that vanishes along certain directions. Thus, as 
  discussed for gapless superconductivity in Section~(\ref{subsec:eilenberger}), 
  these likely will not be useful for sustained operations at GHz frequencies.}
for instance, can vary by two orders of magnitude depending on the angle between the crystal anisotropy axis $\mathbf{c}$ and the applied magnetic field~\cite{tinkham96}. Near zero temperature, the upper critical field of magnesium diboride is about six times larger for $\mathbf{c} \perp \mathbf{B}$ than for $\mathbf{c} \parallel \mathbf{B}$ (see e.g.~\cite{kogan03,budko15}). Here we review Ref.~\cite{liarte16},
which investigates the effects of crystal anisotropy on the superheating field of superconductors, motivated partly by the opportunity of controlling surface orientation in order to achieve higher accelerating fields inside the cavity.

\begin{figure}[!h]
\centering
\includegraphics[width=0.9\linewidth]{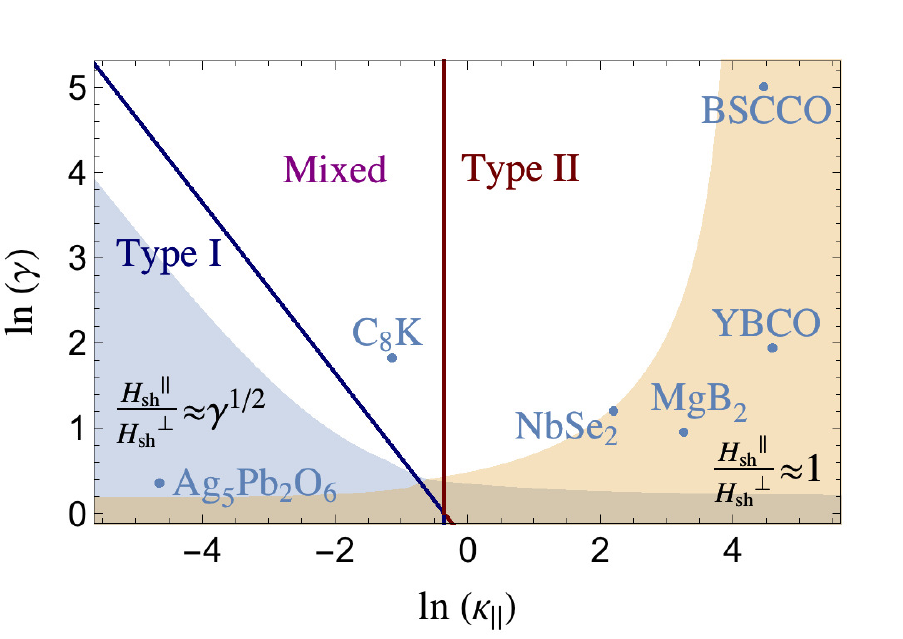}
\caption{Phase diagram of anisotropic superconductors in terms of mass anisotropy ($\gamma = \sqrt{m_c / m_a}$) and GL ($\lambda_a / \xi_a$) parameters (from~\cite{liarte16}). The superconductor is of type-I to the left of the blue line, of type-II to the right of the dark red line, and mixed in between (in the mixed phase, the SC is of type-I for $\mathbf{c}\parallel z$ and of type-II for $\mathbf{c} \perp z$). The blue and yellow regions correspond to the asymptotic solutions ${H_{\mathrm{sh}}}^\parallel / {H_{\mathrm{sh}}}^\perp \approx \gamma^{1/2}$ and ${H_{\mathrm{sh}}}^\parallel / {H_{\mathrm{sh}}}^\perp \approx 1$, respectively (within 10\% accuracy). Note that the superheating field of MgB${}_2$ is nearly isotropic near $T=T_c$.
\label{fig:anisotropyDiagram}}
\end{figure}

Near the critical temperature, for the anisotropy axis $\mathbf{c}$ aligned with one of the Cartesian directions, the anisotropic formulation of Ginzburg-Landau theory~\cite{ginzburg52,caroli66,gorkov64,tilley65} provides a clean approach to study the anisotropy of the superheating field. We can use a change of coordinates and rescaling of the vector potential to turn the anisotropic GL free energy onto isotropic form, and then use previous results from Ref.~\cite{transtrum11} to calculate the superheating field anisotropy of several materials. We find that:
\begin{eqnarray}
H_{\mathrm{sh}}^{\mathrm{ani}} =
\left\{
\begin{array}{ll}
H_{\mathrm{sh}} ( \kappa_\parallel ), & \mbox{for $ \mathbf{c} \parallel z$,} \\
H_{\mathrm{sh}} ( \gamma \kappa_\parallel ), & \mbox{for $\mathbf{c} \parallel x$ or $y$,}
\end{array}
\right.
\end{eqnarray}
where the superheating field on the right hand side is the solution of the linear stability analysis for isotropic Fermi surfaces, which we discussed in Section~(\ref{subsec:linearstability}), using $\kappa = \kappa_\parallel$ and $\kappa = \kappa_\perp = \gamma \kappa_\parallel$ for $\mathbf{c}$ parallel and perpendicular to $z$, respectively. Within GL theory, $\gamma = \sqrt{m_c / m_a} = \lambda_c / \lambda_a = \xi_a / \xi_c$, with $m_i$, $\lambda_i$ and $\xi_i$ representing the effective mass, penetration depth, and coherence length along the $i$-th direction, respectively. Since $H_{\mathrm{sh}} \approx 0.75 \, H_c$ goes to a constant for large $\kappa$, we find that the superheating field is nearly isotropic for most high-$\kappa$ unconventional superconductors. On the other hand, $H_{\mathrm{sh}} \approx 0.84 \, H_c \, \kappa^{-1/2}$ for small-$\kappa$ type-I superconductors, resulting in an anisotropy of about $\gamma^{1/2}$ when $\kappa_\parallel \gamma$ is small. Figure~(\ref{fig:anisotropyDiagram}) displays a phase diagram in terms of $\kappa_\parallel$ and $\gamma$, showing the region where GL theory predicts type-I (left of the blue line), type-II (right of dark red line) and mixed (in between dark red and blue lines) superconductivity, and the regions where each asymptotic solution is expected. Note, in particular, that ${H_{\mathrm{sh}}}^\parallel / {H_{\mathrm{sh}}}^\perp \approx 1$ for MgB${}_2$. This result is valid only very near $T_c$, where the anisotropies in $\lambda$ and $\xi$ are equivalent. In the next paragraph we will use results from a two-gap BCS theory to estimate the superheating field anisotropy of MgB${}_2$ at lower temperatures.

Theoretical and experimental studies indicate that the assumption $\lambda_c / \lambda_a = \xi_a / \xi_c$ (vortex and vortex core have identical shapes within GL theory) is strongly violated for low-temperature MgB${}_2$, thus suggesting the use of two parameters to describe crystal anisotropy, namely $\gamma_{\lambda} = \lambda_c / \lambda_a$ and $\gamma_\xi = \xi_a / \xi_c$. Also, $\gamma_\lambda$ and $\gamma_\xi$ exhibit different temperature dependences, with $\gamma_\lambda$ decreasing and $\gamma_\xi$ increasing for decreasing temperature, respectively. Calculations from Ref.~\cite{kogan03} using a two-gap BCS model suggest that $\gamma_\lambda$ and $\gamma_\xi$ become equal only at $T_c$; near zero temperature, $\gamma_\xi \approx 6$ whereas $\gamma_\lambda \approx 1$, agreeing with some~\cite{cubitt03a,cubitt03b,budko15}, but not all (See Ref.~\cite{kogan03} and references therein) experimental estimates.

\begin{figure*}[!ht]
\centering
(a) 
\includegraphics[width=0.3\linewidth]{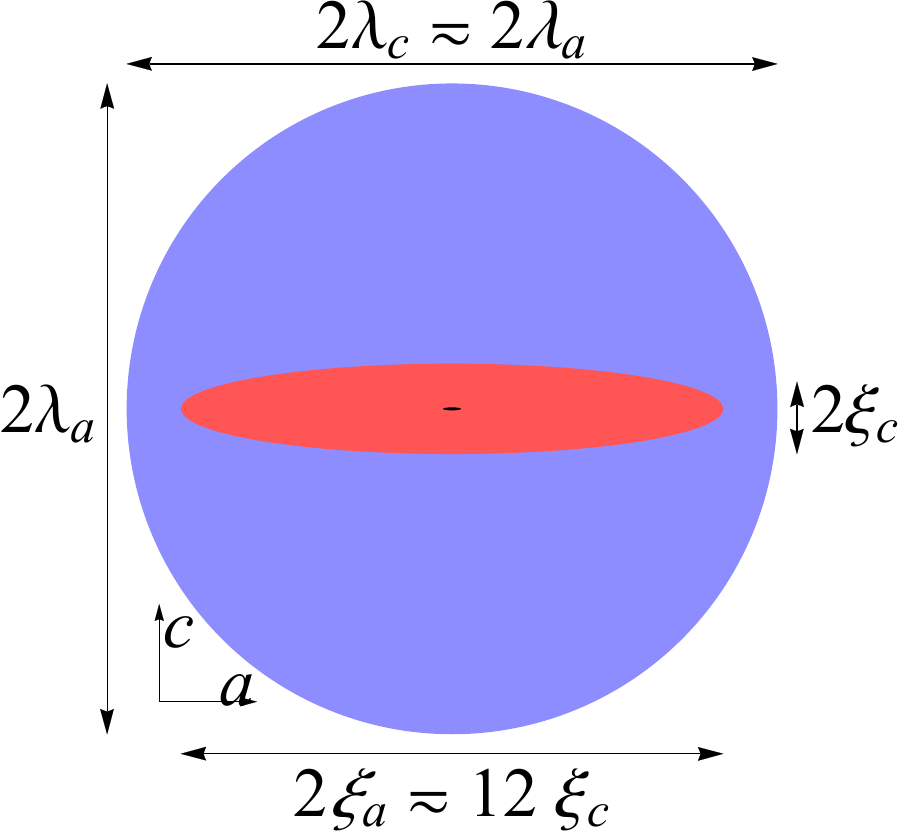}
(b) 
\includegraphics[width=0.6\linewidth]{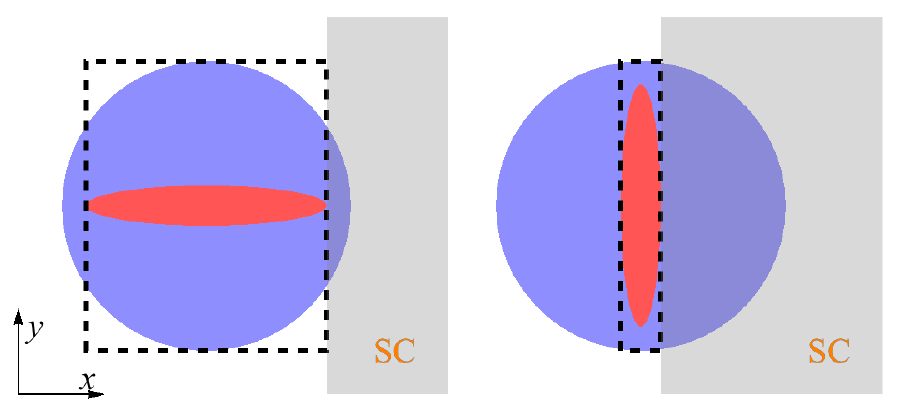}
\caption{From Ref.~\cite{liarte16}. (a) Illustrating vortex (blue disk) and vortex core (red disk) of zero-temperature MgB${}_2$ in the $ac$ plane, with the external magnetic field parallel to the normal of the plane of the figure. We have drawn $\xi_a$ about 30 times larger with respect to $\lambda_a$, so that the core becomes discernible; in the correct scale, the vortex core occupies the tiny black region in the middle of the figure. Notice that vortex and vortex core have identical shapes within GL theory. 
(b) To estimate the superheating field, we calculate the work to push a vortex core into the superconductor, thus destroying the Meissner state. The very different area of the black dashed boxes for $\mathbf{c} \parallel y$ (left) and $\mathbf{c}\parallel x$ lead to substantial anisotropy of the superheating field for low-temperature MgB${}_2$. \label{fig:mgb2Vortices}}
\end{figure*}

We can use our simple estimates of Section~(\ref{subsec:simpleArgument}) to 
make a qualitative prediction for the resulting anisotropy 
$H_{\mathrm{sh}}^{c\perp y} / H_{\mathrm{sh}}^{c\parallel y}$ in the 
superheating field, when $\gamma_\xi \ne \gamma_\lambda$ deviates from the
single-band GL prediction.
Now the anisotropic shape of the vortex and vortex core plays an important role (see Fig.~(\ref{fig:mgb2Vortices})a). When $\mathbf{c}$ is in the $xy$ plane, as in Fig.~(\ref{fig:mgb2Vortices})b, for instance, the superheating field is estimated from the work performed to push the black-dashed ``box'' into the superconductor, which can considerably vary from $\mathbf{c} \parallel y$ (left) to $\mathbf{c} \parallel x$ (right). 
This leads to an estimate 
$H_{\mathrm{sh}}^{c\perp y} / H_{\mathrm{sh}}^{c\parallel y} \approx \gamma_\xi / \gamma_\lambda$.
A second estimate generalizes the Bean and Livingston argument
of the longer-range vortex attraction to incorporate anisotropy, and leads to
a slightly different result: $H_{\mathrm{sh}}^{c\parallel x} /
H_{\mathrm{sh}}^{c\perp x} \approx \gamma_\xi / \gamma_\lambda$.
Yet a third calculation, which we term ``Extended GL'', yields an almost isotropic
result, and is based on a direct linear stability analysis of the anisotropic GL free
energy (see Eq.~(7) of Ref.~\cite{liarte16}) assuming \emph{unconstrained}
$\lambda$'s and $\xi$'s. Table~\ref{tab:mgb2Tab} summarizes
our estimates of $\Hsh$ for the three geometries, using experimental
values for $H_c$ and $\kappa$ for MgB$_2$. Note that we correct numerical
discrepancies of our first estimates in the second row of the table: ``1st
(corrected)''. The last column shows the maximum superheating field
anisotropy according to each method. Most of the values of $\Hsh$ are
as low as $\Hsh \approx 0.24$T for Nb~\cite{padamsee09}. We discuss the origin of these
disparate predictions further in Ref.~\cite{liarte16}.

\begin{table}[h]
\centering
\begin{tabular}{ | l | c | c | c | c | }
\hline
\multirow{2}{*}{Approach}  & \multicolumn{3}{c |}{ $H_{\text{sh}}$ ( Tesla ) } & \multirow{2}{*}{Max. Anis.} \\ \hhline{~---~}
 & $c \parallel x$ & $c \parallel y$ & $c \parallel z$ & \\ \hline
1st estimate & $0.04$ & $0.006$ & $0.04$ & $\sim 6$ \\
1st (corrected) & $0.2$ & $0.03$ & $0.2$ & $\sim 6$ \\
2nd estimate (B \& L) & $1.13$ & $0.18$ & $0.18$ & $\sim 6$ \\
``Extended GL'' & $0.21$ & $0.22$ & $0.22$ & $\sim 1$ \\
\hline
\end{tabular}
\footnotesize
\caption{Estimates of the superheating field and maximum anisotropy of low-temperature MgB$_2$ for three geometries.}
\label{tab:mgb2Tab}
\end{table}

Our GL arguments for the superheating field anisotropy can be trusted near $T_c$: at large $\kappa$ the superheating field anisotropy 
is not a reason to control surface orientation. Our arguments at lower
temperature and for multi-band superconductors are more speculative. The vortex core shape will surely change for $x \sim \xi$
due to the boundary conditions at the surface; the anisotropy in the long-range
attraction in multi-band materials may be different from that of a simple
anisotropic GL approach.
It will be important to apply linear stability analysis to more sophisticated theories, such as multi-gap BCS or strong-coupling Eliashberg theory, especially in the face of the conflicting results shown in Table~\ref{tab:mgb2Tab}.

\subsection{Disorder and vortex nucleation.}
\label{subsec:disorder}

Niobium RF cavities are routinely operated in the metastable regime, at fields
$H_{c1} < H < \Hsh$ above the field $H_{c1}$ where vortices in equilibrium
would penetrate into the superconductor (and dissipate roughly the same
energy as in a normal metal). Table~(\ref{tab:MaterialsParameters})
in Section~(\ref{subsec:thermalFluctuations}) gives
$H_{c1}$ and $\Hsh$ for other candidate materials. For niobium
this metastable regime gives us an important factor of $\sim1.6$
in field. Running in the metastable regime is crucial for utility with the higher temperature
superconductors, whose $H_{c1}$ equilibrium fields are much lower 
than the operating fields for current Nb cavities
(Table~(\ref{tab:MaterialsParameters})). 

It took many years of experimentation to raise operating fields of the niobium
cavities to approach near to their fundamental limits. Will the new, more complex materials
be fundamentally more challenging to optimize? Our preliminary experimental
cavities using Nb$_3$Sn appear already to be operating above
$H_{c1}$~\cite{posen15b}, but are not yet delivering anywhere near to the theoretically predicted
superheating field. Just as we have been exploring the fundamental theoretical
limits to the fields for ideal surfaces, in this section we explore the
fundamental theoretical challenges in minimizing the effects of dirt,
flaws, and defects in lowering the barriers to vortex entry.

What kind of flaw or disorder fluctuation would be needed to allow vortices
to enter at fields substantially lower than the superheating field? How
big a damage region is needed to bypass the surface barrier to vortex entry?
Damage will significantly affect the superconducting properties if the
flaw or fluctuation has a characteristic length of order the coherence
length $\xi$. Since the proposed candidate materials for next generation SRF 
cavities have shorter coherence lengths than niobium 
(Table~(\ref{tab:MaterialsParameters})), this potentially
could imply that these new materials are more susceptible
to defects and dirt.

\begin{figure}[!h]
\centering
\includegraphics[width=0.7\linewidth]{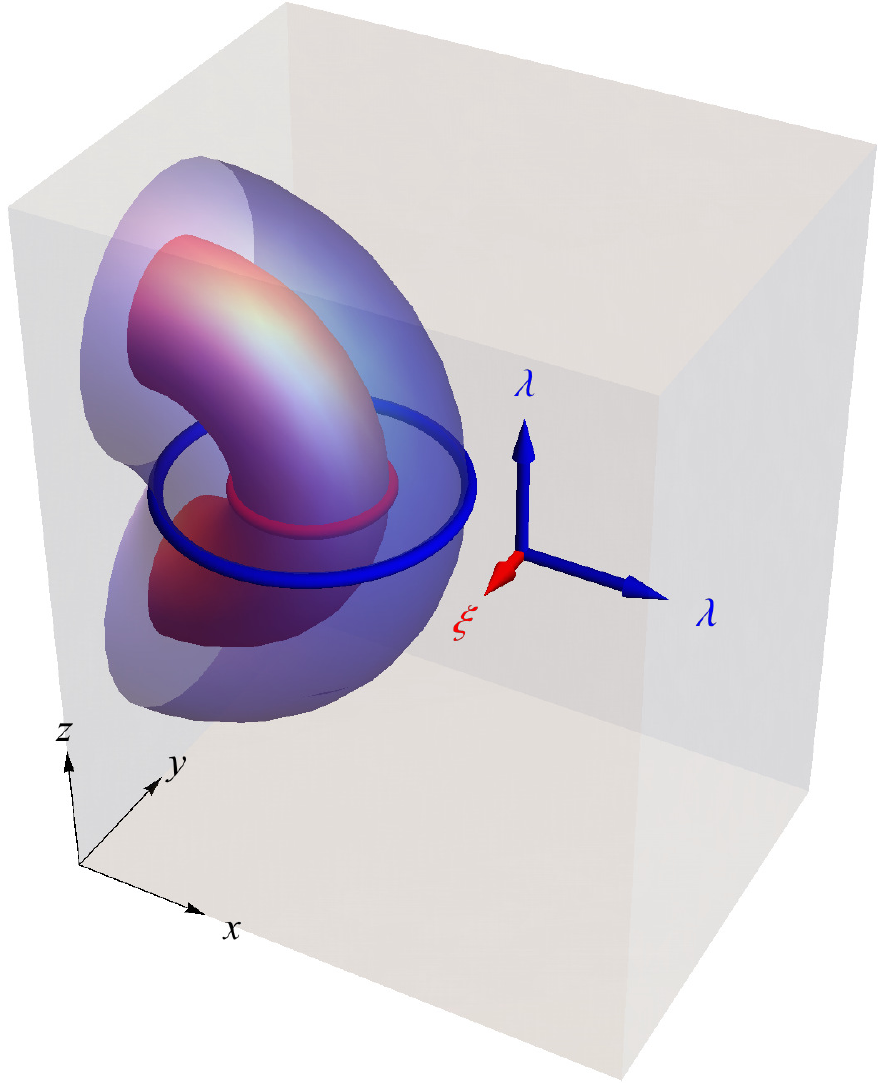}
\caption{{\bf Flux tube nucleation} allowing the penetration of a
single vortex core into the superconductor occupying the half space $x>0$. 
The nucleation barrier at zero disorder can be estimated by computing
the energy of this loop, plausibly a semicircular loop of radius $R$,
and subtracting the magnetic work done by the pressure due to the external
field $H$, where $H_\mathrm{c1} < H < H_\mathrm{sh}$. 
This figure illustrates the nucleation barrier for low fields near $H_{c1}$;
at higher fields the radius becomes comparable to $\xi$. The boundary
conditions at the surface of the superconductor lead to an attractive
force on the vortex as discussed in Section~(\ref{subsec:simpleArgument}),
in addition to the curvature energy of the vortex loop (ignored here).
We can estimate the 
disorder needed to nucleate at a field $H$ by calculating the damage needed
to lower this barrier to zero as the radius $R$ grows.}
\label{fig:fluxTube}
\end{figure}

Figure~(\ref{fig:fluxTube}) shows a cartoon of a vortex loop entering
a superconductor. Based on the discussion at the end of 
Section~(\ref{subsec:simpleArgument}) and the caption of
Fig.~(\ref{fig:fluxTube}), at external fields far from $H_\mathrm{c1}$ and
$H_\mathrm{sh}$, the energy of the vortex loop will grow in the 
absence of disorder until it reaches a critical radius $R_c$,
at which point the energy will again decrease. This critical radius and
the needed damage zone will get smaller as the field $H$ grows, vanishing
at $H=\Hsh$. 

\begin{figure}[h]
(a) \par\smallskip
\centering
\includegraphics[width=0.9\linewidth]{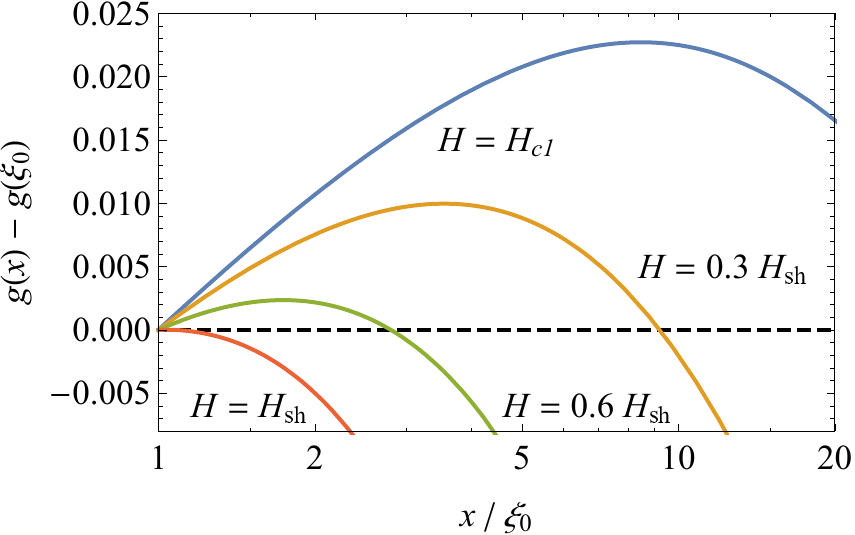}
\\
(b) \par\smallskip
\centering
\includegraphics[width=0.9\linewidth]{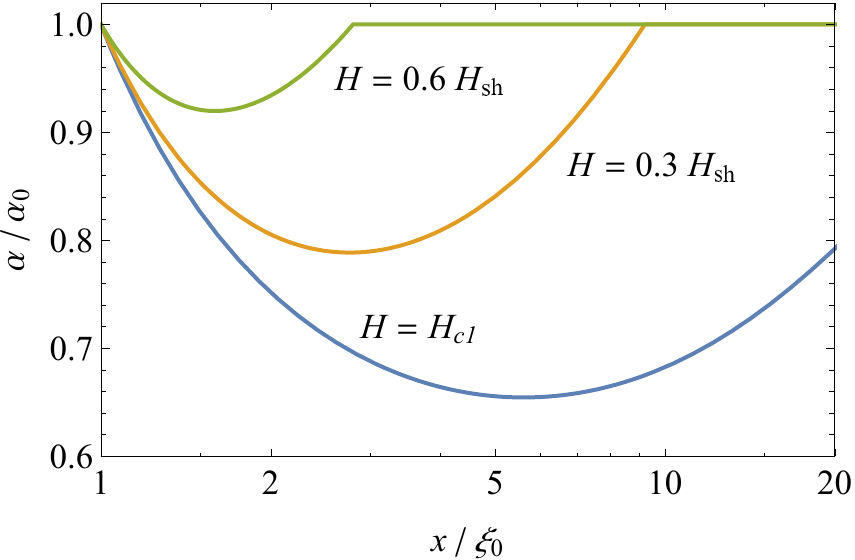}
\caption{
(a)~Unitless Gibbs free energy (Eq.~(\ref{eq:BLG2})) to push a straight
vortex line from a depth $\xi$ to depth $x$ into a superconductor like
Nb$_3$Sn with $\kappa=26.4$, for several values
of the magnetic field. The superheating
field can be estimated in the large $\kappa$ limit from the condition
$G^\prime(\xi)=0$, characterizing the vanishing of the surface energy
barrier at $x=\xi$; Bean and Livingston's estimate gives 
$h_{\mathbf{sh}}=1/2$ so $\Hsh = H_c/\sqrt{2} \approx 0.71 H_c$, comparable
to the correct large-$\kappa$ limit.
Note that the peak in the barrier is at 
$x_c \sim \xi \Hsh/H$; near $H_{c1}$ it is roughly
$\lambda \approx \kappa \xi/\log \kappa = \lambda/\log \kappa \approx \lambda$,
but in the interesting region near $\Hsh$ it is near the coherence length
$\xi$.  (b)~The spatially-dependent critical
temperature shift $\alpha=\alpha(x)$, needed to
flatten the energy barrier and allow for the penetration of vortices, in our particular model 
with $\kappa$ for Nb$_3$Sn.
This is shown for several values of $H$ in the interval $[H_{c1},\Hsh)$. Here $H=0.6 H_c$ would
duplicate the maximum possible superheating field for niobium.
\label{fig:energyBarrier}}
\end{figure}

The energy per unit length
of the vortex loop will have two contributions -- a curvature energy and 
an attractive energy between the vortex and the surface. The latter can
be estimated from the attraction of a straight vortex to the `image vortex' 
needed to set the correct boundary conditions at the surface. This potential
barrier (the major component of the superheating field) was estimated
by Bean and Livingston for 
high $\kappa$ type-II superconductors~\cite{bean64}. The unitless Gibbs
free energy per unit length $4 \, \pi \, G/(\sqrt{2} \, {H_c} \, \Phi_0)$
of a straight vortex flux line a depth $x$ inside a superconductor
with external field $H$ can be written in the
(London) large-$\kappa$ limit as~\cite{huebener01}:
\begin{align}
\label{eq:BLG1}
G = \frac{\Phi_0}{4\pi} \left(H (e^{-x/\lambda} -1) 
  - \frac{1}{2} \frac{\Phi_0}{2 \pi \lambda^2} K_0 (2 x/\lambda) 
  + H_{c1}\right) \\
\frac{ 4 \, \pi \, G }{ \sqrt{2} \, {H_c} \, \Phi_0} = g(x) = h \left(e^{-x/\lambda} -1\right) - \frac{K_0 (2\,x / \lambda)}{2\,\kappa} + \frac{\ln \kappa}{2\,\kappa}, \label{eq:BLG2}
\end{align}
where $h = H/(\sqrt{2}H_c)$, and $K_\nu$ denotes the modified Bessel
function of imaginary argument~\cite{gradschteyn07}, and for now
$\kappa$ and $\lambda$ are the Ginzburg-Landau parameter and penetration length of the pure material, respectively.
The first term is
a magnetic pressure, the second term is the interaction 
with the `image vortex' that imposes the correct boundary conditions, and
the third term is the energy per unit length of a vortex deep in the 
superconductor. We can estimate $R_c(H)$
by setting the derivative $dG/dx = 0$ in Eq.~(\ref{eq:BLG1}) and expanding
the Bessel function for small arguments, leading to 
$R_c(H) \sim \xi \Hsh/H$. At the lowest field for vortex penetration $H_{c1}$ this expansion is unreliable; however, since $H_{c1} = \Hsh \log(\kappa) / \kappa$~\cite{huebener01}, the resulting estimate $R_c(H_{c1}) \approx \xi \kappa/\log(\kappa) = \lambda/\log(\kappa)$ is still quite good, as shown in Figure~(\ref{fig:energyBarrier}a). But the new materials of interest have lower critical fields $H_{c1}$ too small to be useful; we must run at fields $H$ comparable
to $\Hsh$. Near $\Hsh$, $R_c \approx \xi$, again as shown in Fig.~(\ref{fig:energyBarrier}).
For disorder or defects to remove this energy barrier, they will thus 
necessarily have to strongly affect a region of volume
$\sim \xi R_c^2 \sim \xi^3$.

To make this more quantitative, one needs to identify and model the dominant
mechanism for vortex nucleation. If the characteristic defect size is large
compared to $\xi$ (e.g., nucleation on grain boundaries or inclusions
of competing phases), one must model and control these individual defects.
Clean grain boundaries are usually atomistically sharp (much thinner than
$\xi$) and hence do not significantly decrease the local superconducting 
properties; indeed, studies of hot spots in large grain niobium cavities show no
correlation with grain boundaries~\cite{eremeev06}, and using single
crystals to avoid grain boundaries has not improved
performance~\cite{kneisel05a,kneisel05b}. But in more complex materials, grain
boundaries could be more disordered, thicker, or contaminated by impurities, 
and a grain boundary or grain boundary intersection with the correct
orientation with respect to the surface could provide a route to entry.
The effect of surface roughness on Bean and Livingston's surface
barrier has been studied in Ref.~\cite{bass96}. Kubo has used the
London model to investigate the effects of nano-scale surface topography on
the superheating field~\cite{kubo15}. 
Perhaps most dangerous could be inclusions of metallic or poorly 
superconducting second phases, or irregularities in the surface morphology.


If the characteristic defect size is small compared to $\xi$, and if the
defects are uncorrelated in position, then the fluctuations in
regions of order $\xi^3$ can be quantitatively estimated to linear order
using the central limit theorem.
This leads to Gaussian random fluctuations in the superconducting properties.
For example,
for alloys and doped crystals there are natural concentration fluctuations
that will locally change the superconducting transition temperature, 
coherence length, condensation energy, and other properties. 
This is the traditional theoretical framework for field-theoretic 
calculations of the effects of disorder. 

Let us hypothesize a system where the critical temperature is decreased due to
disorder. In the context of Ginzburg-Landau theory for a homogeneous system,
a change in the critical temperature yields a change in the coefficient $\alpha=\alpha(x)$
of $\psi^2$, where $\psi$ is the superconductor order parameter~\cite{tinkham96}.
The probability of a fluctuation in $\alpha(\mathbf{x})$ away
from its pure value $\alpha_0$ would be proportional to
\begin{equation}
\label{eq:Pofalpha}
\Pi\{ \alpha ({\mathbf{x}})\} \propto
    \exp \left( 
            -\int (\alpha ({\mathbf{x}}) / \alpha_0 -1)^2 / (2 \sigma^2) d^3 {\mathbf{x}}
        \right),
\end{equation}
where $\sigma$ is a material-dependent constant that encapsulates the
likelihood that the dirt in the material will cause a given fractional
change $\alpha/\alpha_0$ in the critical temperature. The constant $\sigma$ will
become larger either if there are bigger concentration fluctuations or
if the material is particularly sensitive to dirt. In principle, we
should now calculate the most probable three-dimensional profile
$\alpha(\mathbf{x})$ needed to flatten the energy barrier and allow vortices in
at a lowered field $H<\Hsh$, and then use $\Pi\{ \alpha ({\mathbf{x}})\}$ in
Eq.~(\ref{eq:Pofalpha}) to estimate the probability per unit surface area
$P(H/\Hsh)=\Pi\{\alpha(\mathbf{x})\}$ of vortex penetration.

Rather than doing this full variational calculation, we
build on the Bean-Livingston model of Equation~(\ref{eq:BLG1}).
In GL theory, the characteristic lengths scale as $\lambda \sim \alpha^{-1}$
and $\xi \sim \alpha^{-1/2}$. Hence we distinguish $\lambda_0$, $\xi_0$ and
$\kappa_0$ for the pure material from $\lambda(x)=\lambda_0/a$, $\xi(x)=\xi_0/\sqrt{a}$ and
$\kappa(x) = \lambda(x)/\xi(x)=\kappa_0/\sqrt{a}$ for the damaged region, where $a=\alpha/\alpha_0$.

\begin{figure}[!h]
\centering
\includegraphics[width=0.9\linewidth]{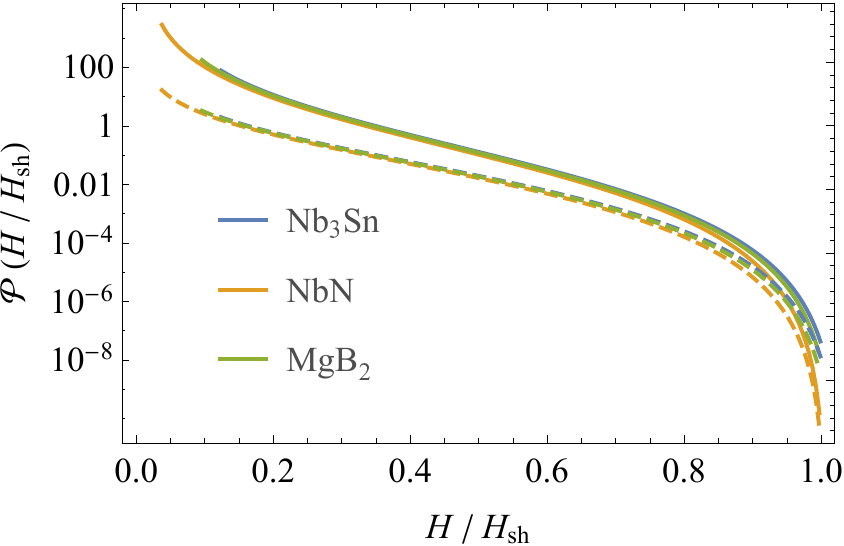}
\caption{{\bf Relative logarithmic reliability
${\cal P} = -(2\sigma^2 /\xi_0^3)\log(P(H/\Hsh))$ of vortex nucleation}, in a
simple model of Gaussian random disorder, for the $\kappa$ values of the
three candidate superconductors. Solid curves are ${\cal P}_{3D}$
for a semicircular vortex barrier model 
(Fig.~(\ref{fig:fluxTube}), Eq.~(\ref{eq:P3D})); dashed
curves are $(d/\xi_0) {\cal P}_{2D}$ for pancake vortex nucleation in
a 2D superconducting layer of thickness $d$ (Section~(\ref{sec:laminates})).
\label{fig:logPvsH}}
\end{figure}

What is the minimum amount of dirt that is necessary to reduce the
superheating field to a certain value? For instance, how much dirt would
it take to reduce $\Hsh$ for Nb$_3$Sn (estimated at 0.42 T in 
Table~(\ref{tab:MaterialsParameters})) to  $H = 0.25$ T ($\Hsh$ for niobium), 
a factor $H/\Hsh \sim 0.6$? One would need enough dirt `flatten' the
surface barrier between%
  \footnote{Bean and Livingston measure the barrier starting at $x=\xi$, 
  below which London theory is unreliable.}
$\xi_0$ and $R_c(H) \approx \xi_0 \Hsh/H = 5 \xi_0/3$ along the $x$ direction (thus
allowing for vortex penetration), as shown in the dashed line of
Fig.~(\ref{fig:energyBarrier}a). In general, we are interested in finding
an $x$-dependent parameter $\alpha=\alpha (x)$ that flattens the energy
barrier from $x=\xi_0$ to $x=x_f$, where $x_f>\xi_0$, and is defined by
$G(x_f) = G(\xi_0)$. The solution for $\alpha(x)$ is then found from the
equation $G(x) = G(\xi_0)$ for $\xi_0<x<x_f$, and $\alpha (x) = \alpha_0$ for
$x>x_f$, where in the left and right hand sides we use $\{\lambda(x), \xi(x)\}$ and
$\{\lambda_0, \xi_0\}$ in Eq.~\eqref{eq:BLG1}, respectively.

Note that we are making a rough approximation here. The magnetic fields and 
supercurrents surrounding the vortex line will see a spatially varying critical
temperature $\alpha(\mathbf{x})$ whenever it is far from the surface, and
properly measuring its energy and thus the surface attraction should include the
resulting shift in energy. The depths $x$ of importance to us are of order the 
coherence length $\xi$, and thus these long distance fields and currents are largely cancelled
by the image vortex a distance $2x$ away. The vortex will see a depth-dependent disorder,
but its energy will be qualitatively well described by our model in the region $H\sim\Hsh$.

Figure~(\ref{fig:energyBarrier}) shows $\alpha / \alpha_0$ as a
function of $x/\xi$ for several values of $H$ in the interval
$[H_{c1},\Hsh)$. Apart from an overall constant given by the normalization
of the Gaussian, the negative logarithm of the probability of this fluctuation
as a function of the lowered entry field $H$ is
\begin{align} 
-\log(P(H/\Hsh)) &= \Pi\{\alpha(\mathbf{x})\} \\
&= \int (\alpha(\mathbf{x})/\alpha_0-1)^2/(2\sigma^2)d^3\mathbf{x} \\
&= \frac{\xi_0^3}{2\sigma^2} 
		\int  (\alpha(\xi_0 \mathbf{u})/\alpha_0-1)^2 d^3\mathbf{u} \\
&= \frac{\xi_0^3}{2\sigma^2} 
		{\cal P}(H/\Hsh),
\end{align}
a measure of the relative logarithmic reliability of the superconductor to 
disorder-induced nucleation. Here we pull out the volume $\xi_0^3$ of
the damage zone by changing variables to $\mathbf{u} = \mathbf{x}/\xi_0$.
In a three-dimensional system with a semicircular vortex nucleation
approximation, we can use our Bean-Livingston style methods to approximate
this as a one-dimensional integral
\begin{equation}
\label{eq:P3D}
{\cal{P}}_{3D}(H/\Hsh) = \int \pi u (\alpha(\xi_0 u)/\alpha_0-1)^2 du.
\end{equation}
For a vortex pancake nucleation event for a thin SIS film of thickness $d$
(discussed in Section~(\ref{sec:laminates})), we find
\begin{equation}
\label{eq:P2D}
{\cal P}_{2D}(H/\Hsh) = (d/\xi_0) \int (\alpha(\xi_0 u)/\alpha_0-1)^2 du
\end{equation}
(see Fig~(\ref{fig:logPvsH})).

Clearly, the relative reliability decreases rapidly as $H$ approaches $\Hsh$,
by many orders of magnitude in this model calculation.
The high-$\kappa$ calculation of Bean and Livingston cannot be simply
extrapolated to niobium, but there is no reason to doubt that a similar
sensitivity of the barrier to $H/\Hsh$ is expected. Nonetheless, niobium
cavities are used in planned applications at $0.7\Hsh$~\cite{ILCtdr,geng11},
suggesting realistic values of disorder are tolerable in niobium.
Indeed, the dependence of the barrier on $H/\Hsh$
is much stronger than its dependence on $\kappa$ or $\xi$. This
suggests, examining Figure~(\ref{fig:logPvsH}), that the
factor of five to ten change in $\xi_0$ with the new superconductors
may not be so dangerous. The resulting two to three orders of magnitude
smaller volume for the critical damage zone at fixed field, it would seem,
could be remedied by working not at $0.8\Hsh$
but at perhaps $0.6\Hsh$ (Figure~(\ref{fig:logPvsH})). Manufacturing high-quality
cavities from these new materials may be challenging.
What our calculation can provide is reassurance
that these materials should not be avoided because of their shorter coherence
lengths.

%% file: Sec/Laminates.tex
\section{Laminates and vortex penetration}
\label{sec:laminates}



In recent years, much effort in superconducting RF has been devoted to exploring
single or multiple thin films -- laminated structures hopefully tunable to optimize performance. This section is devoted to exploring possible advantages
to such laminates. The work in this section relies heavily on extensive
discussions and consultation with Alex Gurevich, whose work prompted most
of the calculations presented.

In practical terms, two of the candidate materials (Nb$_3$Sn and
NbN) can be grown by deposition on Nb surfaces, so fabricating a 
surface layer onto a Nb cavity leverages existing expertise. Gurevich
points out~\cite{Gurevich2013} that thermal conductivities of new candidate materials are often small; since the heat generated by the surface residual 
resistance at the surface must be conducted through the cavity, keeping the
thickness of these new materials small can improve performance. (For 
Nb$_3$Sn, recent surface resistances have been small enough, at least at low
fields, that thickness may not be an issue.) Gurevich
has also proposed {\cite{Gurevich2006} separating one or more
superconducting layers by insulating
layers (a SIS geometry). Calculations show~\cite{posen15} that laminates
do not substantially improve the theoretical maximum superheating field in AC applications beyond 
that of pure materials (or thick layers) for the film-insulator-bulk structure,%
    \footnote{\label{foot:IsolatedLayer}
    A free-standing superconducting layer (or a layer
    surrounded by insulators) with thickness small compared to the magnetic
    penetration depth $\lambda$ can have an enormous superheating field
    (since it can remain superconducting without paying most of the cost of
    expelling the flux).
    In the accelerator community, there is widespread focus
    on raising this `$H_{c1}$' for the superconducting 
    film~\cite{feliciano16,lamura09}
    -- defined, somewhat unphysically~\cite{Hc1WrongRef} as the minimum field
    needed for a vortex to be stable parallel to and inside the film.
    But such an in-film stable vortex configuration demands magnetic flux on 
    {\em both sides} of the film. In a GHz AC application, pushing the 
    flux through the film twice per cycle generates
    unacceptable heating~\cite{Hc1WrongRef}. Besides, any such 
    parallel vortex would be precariously unstable to formation of two vortex
    pancakes. A thin superconducting layer with a large magnetic
    penetration depth atop a lower-\Hsh\ layer with a small penetration
    depth can have modestly higher superheating fields, due to the way the 
    bottom layer modifies the magnetic field penetration.
    }
though adding a thin S$^\prime$ layer on the bulk S superconductor may lead to an enhancement of the energy barrier~\cite{Kubo2016a,Checchin2016}.

Gurevich has suggested that the SIS geometry may have a different advantage --
reducing the impact of flux penetration. Our calculations in 
Section~(\ref{subsec:disordmeander}) suggest that SIS films with thickness
$d$ small compared to the London penetration depth $\lambda$ will be 
more susceptible to vortex penetration than bulk films; the damage zone needed for vortex nucleation at fields below pure $\Hsh$ 
can be thinner by the fraction $\lambda/d$, presumably making them much more likely. Also, one would naively expect
it to be harder to grow low-defect two-layer laminates than depositing
a single layer or preparing a pure surface. Layers thick compared
to the penetration depth would presumably behave similarly to a bulk material;
vortices deeper than $\lambda$ do not `feel' the surface except insofar
as other vortices penetrating the surface push them deeper.

The dynamics after flux penetration will be substantially different
for the SIS geometry than for a simple 3D superconducting surface. In either
case, a flaw may nucleate one to several vortex entries when the field
increases in one direction; some or all may be `pulled back' as the field
shifts to the opposite direction. If the nucleation center flaws are rare
and the vortices do not build up over time, they need not cause local heating
enough to cause a quench. But since the RF cavities operate at GHz frequencies,
and each flaw could (or should) generate multiple vortices per cycle,
potentially billions of vortices could be introduced by a single flaw
if they can escape re-annihilation.

In three dimensions, a vortex penetrating at a point $(y,z)$ on the
surface will grow in the $z$ direction pointing along the field as
it penetrates a depth of order $x\sim\lambda$ (Fig.~(\ref{fig:semiloops}) left).
If multiple vortices enter, they may push and entangle one another; as they interact with disorder
in the material they may exhibit avalanches~\cite{aranson01, aranson05}. During the field reversal, the points where the
vortices exit the material will be forced together along the $z$ direction
(shrinking in length), and new vortices with opposite winding number will
nucleate (potentially annihilating some or all of the old vortices). Even
if this process is incomplete, leaving some tangle of vortex loops, it
may enter a kind of limit cycle. Indeed, many periodically stressed
disordered dynamical systems can enter into limit cycles at low levels
of stress, with a transition to `turbulent' aperiodic behavior at a critical
threshold (colliding colloids in reversing low-Reynolds number flows~\cite{corte08},
plasticity in vortex structures of superconductors~\cite{daroca11,motohashi11,okuma11}, etc).
It is possible that the quench of RF cavities explores precisely this kind of dynamical
phase transition, separating a local hot spot from an invading front of vortices. Apart from these brief speculations, we will not
discuss three-dimensional dislocation dynamics further in this work;
the remainder of this section will focus on the SIS geometry.

In the two-dimensional SIS geometry, a vortex penetration event may end 
with the vortex
trapped in the insulating layer, leaving two 2D vortices penetrating the 
outer superconducting film (Fig.~(\ref{fig:semiloops}) right. See also
footnote~\ref{foot:IsolatedLayer}.)
Such 2D vortices, called {\em pancake vortices},
have been studied in great detail~\cite{huebener01} in the context of high
temperature cuprate superconductors, some of which are well described as 
nearly decoupled 2D superconducting sheets. A vortex pair nucleated 
by a defect at
$(y, z)$ on the surface will separate along the $z$ direction
as the field increases, be buffeted by thermal fluctuations, dirt, defects,
and other vortices as they separate, and then be pulled back along the
$z$ direction as the field reverses.
(Some of the other vortices will be emitted by the same defect, 
once the initial pair departs and the resulting long-range suppression 
of nucleation drops, see Section~\ref{subsec:interacting}.)
In this part we shall explore
Gurevich's suggestion that, even after billions of cycles, this annihilation
should be effective at avoiding vortex escape (presumably preventing
a buildup of vortices which otherwise would lead to a quench).

In Section~(\ref{subsec:impactParameter}), we introduce an ``impact parameter,'' the amount of lateral vortex separation between a vortex-antivortex pair that can be tolerated during a cycle while still expecting them to annihilate, in Section~(\ref{subsec:thermal}), we examine the expected lateral meandering distance expected from pancake vortices in an RF cycle, in Section~(\ref{subsec:disordmeander}), we examine the expected meandering due to disorder, and in Section~(\ref{subsec:interacting}), we briefly consider the effect of vortex-vortex interaction
and the situation of two nearby defects.

\begin{figure}[!h]
\centering
\includegraphics[width=0.9\linewidth]{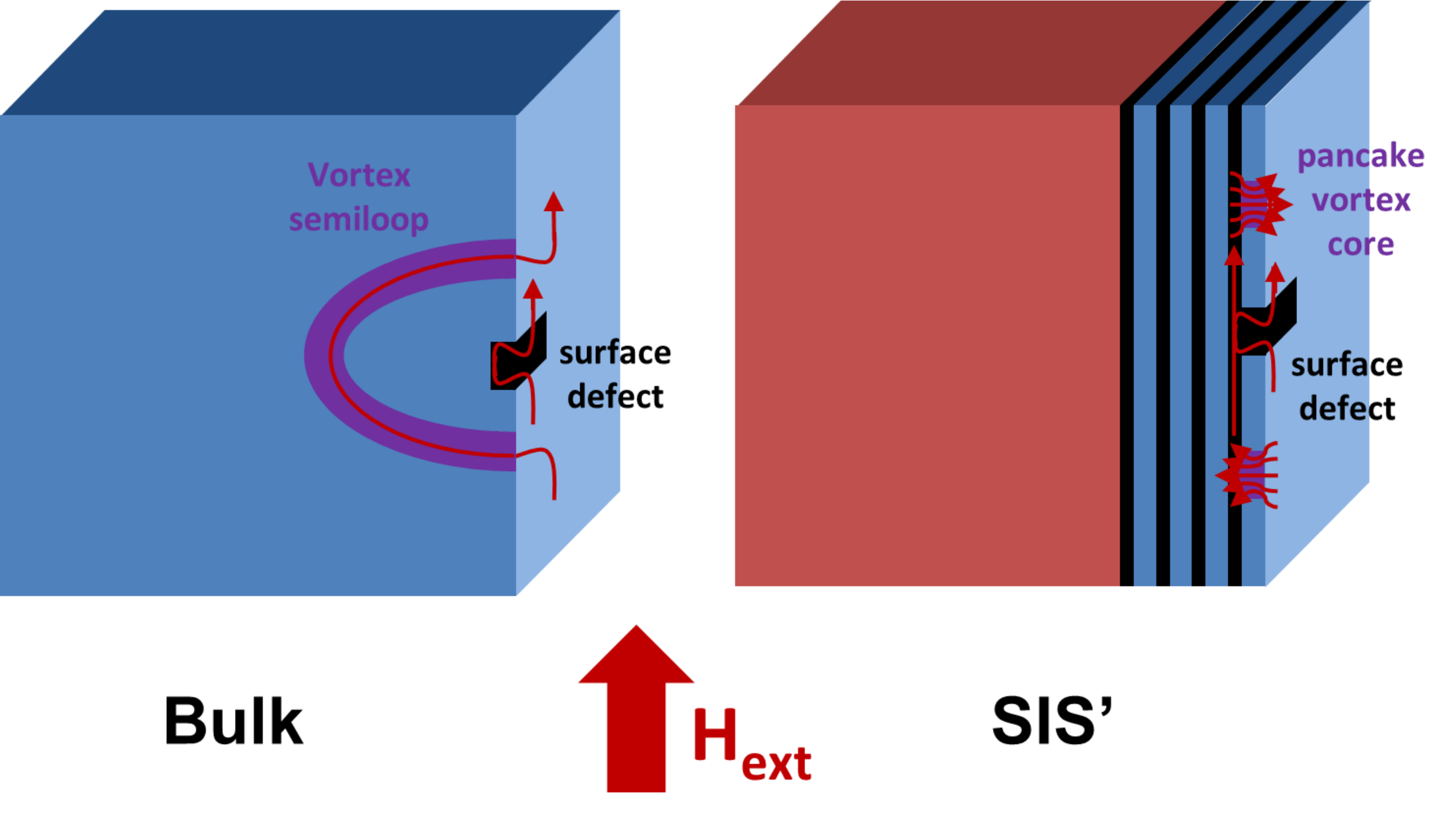}
\caption{Vortices in a bulk superconductor for semiloops (left). Vortices in thin superconducting films separated by insulators form pancakes.\label{fig:semiloops}}
\end{figure}

\subsection{Impact Parameter}
\label{subsec:impactParameter}

How far $\Delta x$ perpendicular to the field must a vortex pair migrate 
before their mutual attraction ceases to be strong enough to annihilate them
at the end of a cycle? Figure~(\ref{fig:separatrix}) shows the trajectories for a pancake pair as they return at the end of a cycle, separated by
different distances $x$ perpendicular to the external magnetic field, using the vortex interaction formulation from Ref. \cite{Clem1991}. There is a {\em separatrix} between trajectories which collide and trajectories which miss each other. We will call the value of $\Delta x$ at this separatrix the {\em impact parameter}, $x_{imp}$. For perhaps credible parameters $d=30$ nm, $\lambda=100$ nm, $\mu_0\Hsh=0.4$ T, $x_{imp} \sim 20$ nm. Simulations were used to evaluate $x_{imp}$ as a function of field, and the results are plotted in Fig.~(\ref{fig:wandercalc}).

\begin{figure}[btp]
\begin{center}
\includegraphics[width=0.9\linewidth,angle=0]{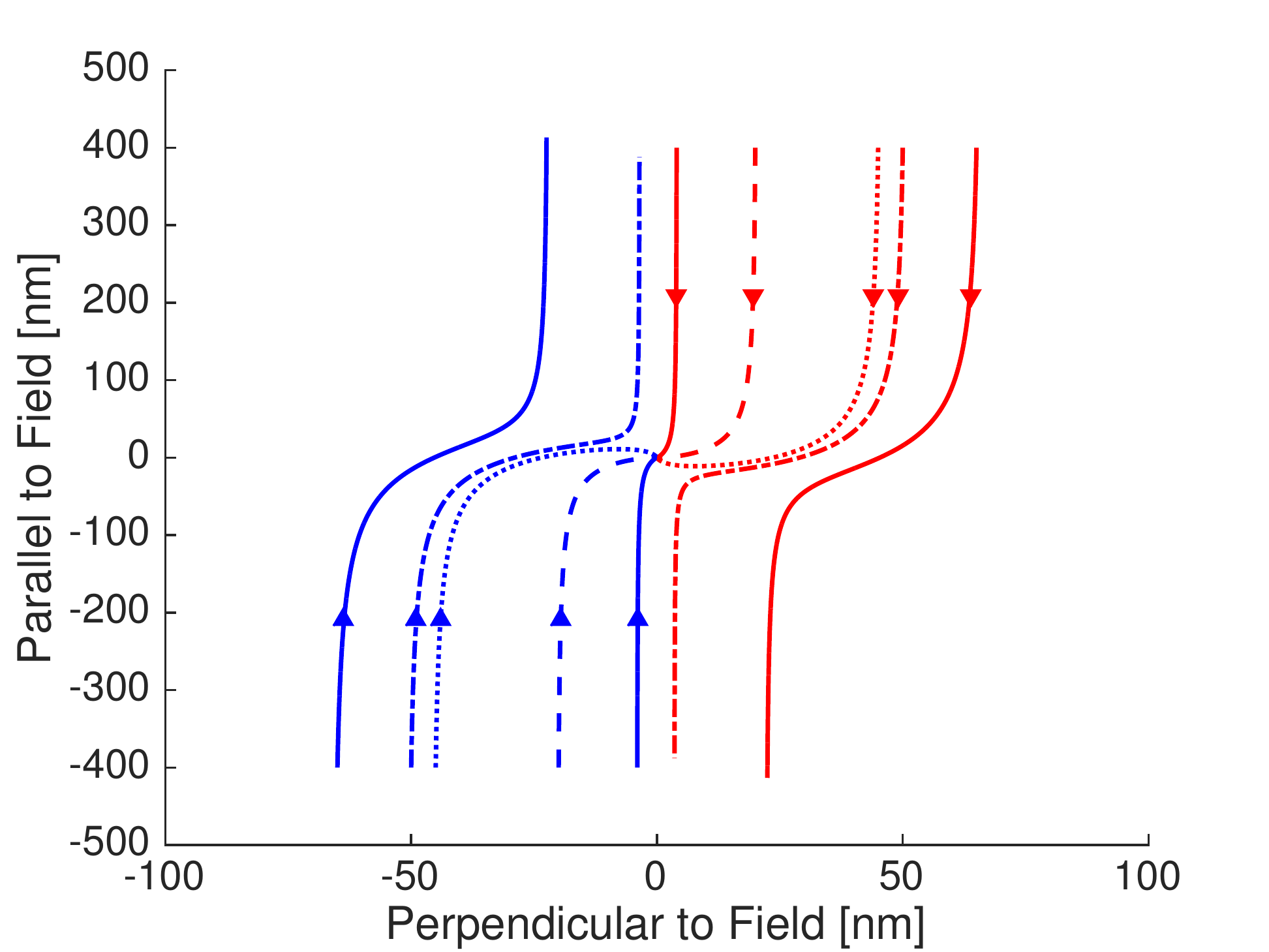}
\end{center}
\caption{For initial separations smaller than the impact parameter $x_{imp}$, the pancakes annihilate (inner solid, dash, dot), but larger than $x_{imp}$, they can wander away (dash-dot, outer solid). $\alpha=0.2$ shown}
\label{fig:separatrix}
\end{figure}

\begin{figure}[btp]
\begin{center}
\includegraphics[width=0.9\linewidth,angle=0]{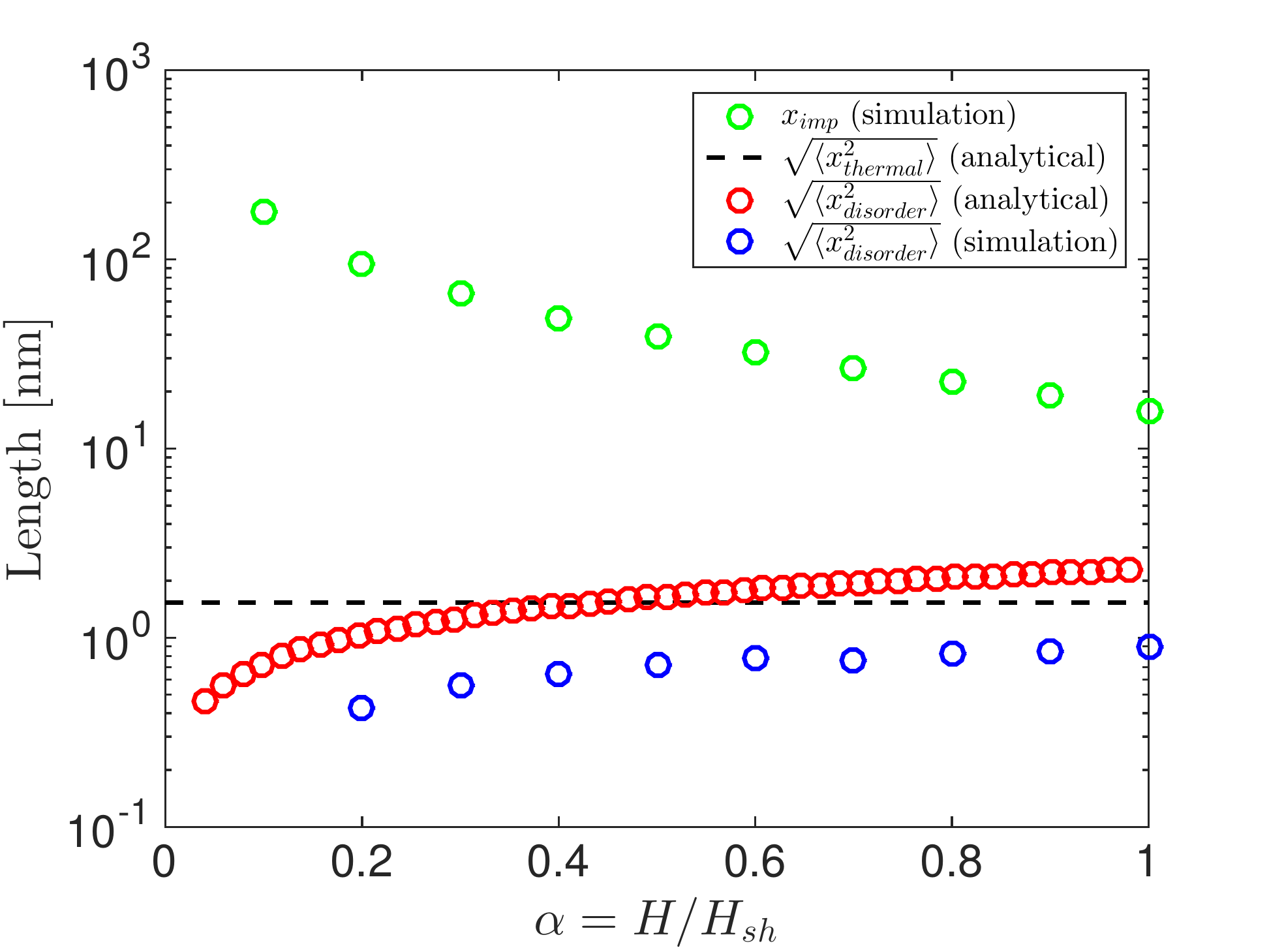}
\end{center}
\caption{This figure compares $x_{imp}$, the maximum lateral separation resulting in impact of a vortex pair, to $\sqrt{\left<{x_{thermal}^2}\right>}$ and $\sqrt{\left<{x_{disorder}^2}\right>}$, the expected meandering distances due to thermal and disorder effects, for realistic parameters given in the text. Note
that the former remains a factor of at least ten larger than the latter,
suggesting that vortex escape by these mechanisms is a $10\sigma$ event.
Thus neither thermal motion nor disorder is dangerous, according to our
estimates, to prevent nucleated pancake vortices from annihilating with extremely high
probability at the end of every cycle.}
\label{fig:wandercalc}
\end{figure}


\subsection{Thermal Meandering}
\label{subsec:thermal}

The motion due to thermal fluctuations can be estimated using the Einstein equation,
\begin{equation}
\left<{x_{thermal}^2}\right>=2Dt
\label{eq:einstein}
\end{equation}
where $x_{thermal}$ is the displacement in time $t$ due to thermal motion and $D$ is a diffusion constant. For one RF cycle at frequency $f$, $t=f^{-1}$. $D\approx\mu k_B T$, where $k_B$ is Boltzmann's constant, $T$ is the temperature and $\mu$ is the mobility of the vortex, given by Bardeen Stephen as $\rho_n/(H_{c2}\,\phi_0\, d)$, where $\rho_n$ is the normal state resistivity, $H_{c2}$ is the upper critical field and $\phi_0$ is the flux quantum. Solving, the wandering due to thermal motion is given by:

\begin{equation}
\sqrt{\left<{x_{thermal}^2}\right>}=\sqrt{\frac{2\,k_B \, T\, \rho_n}{H_{c2} \, d \, f \, \phi_0}}
\label{eq:einstein2}
\end{equation}

For realistic parameters $T=2$ K, $\rho_n=100$ n$\Omega$m; $\mu_0H_{c2}=30$ T, $f=1.3$ GHz, $d=30$ nm, $x_{thermal} = 1.5$ nm, as shown in Fig.~(\ref{fig:wandercalc}). From these results, we can calculate the approximate expected rate of production of vortices that fail to annihilate. One expects that the distribution of final separations will be Gaussian with standard deviation $x_{thermal}$, suggesting that the number of vortices which do not annihilate will be given by the tail of the Gaussian. For example, at $H=0.8\Hsh$, $x_{imp}$ is about 22 nm from Fig.~(\ref{fig:wandercalc}), or about 15 standard deviations, making it extremely unlikely for vortices to escape due to thermal meandering alone. 

\subsection{Disorder Meandering}
\label{subsec:disordmeander}

\begin{figure}[!h]
\centering
\includegraphics[width=0.9\linewidth]{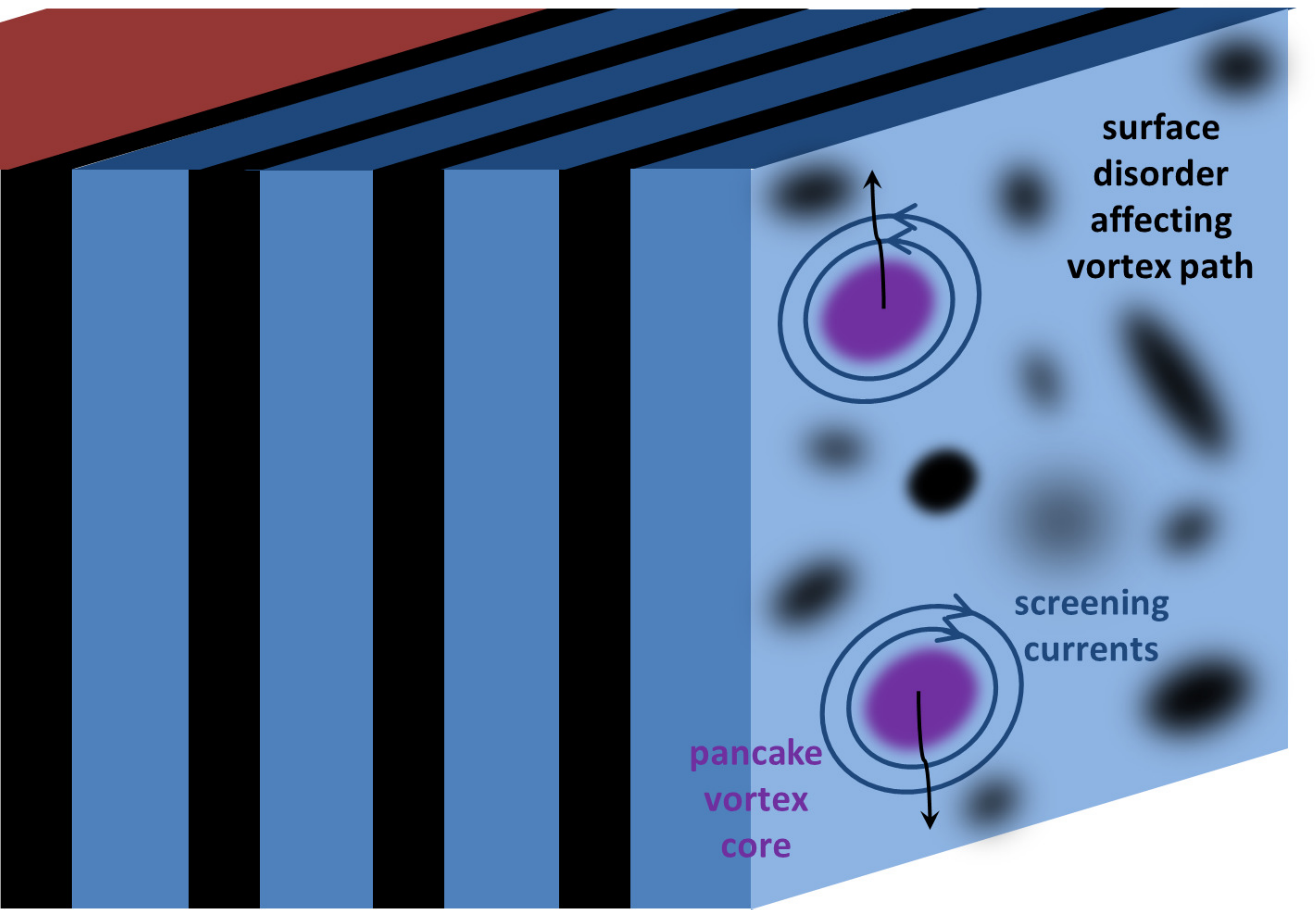}
\caption{Surface disorder may cause pancake vortices to meander away from a nucleation site and build up in a film over many RF cycles. \label{fig:disorder}}
\end{figure}

To calculate the wandering due to surface disorder, illustrated in Fig.~(\ref{fig:disorder}), we consider a single-cell $f$=1.3 GHz niobium SRF cavity with an SIS structure using $d=30$ nm thick Nb$_3$Sn layers. Assume that the topmost S layer has a normally-distributed random array of defects over its surface. For our geometry, we divide the $L\times L$ (where $L\sim$10 cm) surface area of the cavity into $N$ $a\times a$ regions of order the pancake vortex size, where $N=L^2/a^2$. We represent the effect of these defects as lowering the local value of $B_c$ in a given region. Therefore these defects will nucleate vortex penetration, and they will attract pancake vortices in the film. At the worst of the defects, the expected value for $H_c$ is $\alpha H_{c,nominal}$, where $\alpha$ is a constant between 0 and 1. At this defect, vortices penetrate at approximately $H=\alpha \Hsh$. We represent the surface of the cavity with a distribution $\mathcal{H}$ of values for the reduction in the square of $H_c$. For simplicity of analysis, and since the defects are normally distributed, we will use the notation generally used in propagation of random errors.

\begin{equation}
\mathcal{H}=H_{c,nominal}^2(1-|(0\pm\sigma)|)
\label{Bdist}
\end{equation}

Here is $\sigma$ is the variance of the normally distributed $H_c^2$ reduction. We use absolute values because there should be no $H_c$ values higher than the nominal value.

We can find $\sigma$ using our restriction that the expected value for $H_c$ at the worst defect is $\alpha H_{c,nominal}$. To do this, we need to work with a normal distribution of the $H_c^2$ reduction in our N regions $\mathcal{N}(0,\sigma)=[1/(\sigma\sqrt{2\pi})] \exp [ x^2 /  (2\,\sigma^2)]$. First, we need to find $\phi$, the value of $x$ above which lies just one half of one of our N regions (one half because the absolute value effectively doubles the number of samples in our integration).

\begin{equation}
\int_{\phi}^\infty \frac{1}{\sigma\sqrt{2\pi}}e^{-x^2 / (2\, \sigma^2)} \mathrm{d}x=\frac{1}{2}\left[1-\mathrm{erf}\left(\frac{\phi}{\sqrt{2}\sigma}\right)\right]=\frac{1}{2N}
\label{sigmadef1}
\end{equation}

Next, we set the expected value of the distribution in this region to be 1-$\alpha^2$. This sets the expected value of $H_c$ to be $\alpha H_c$ for the most extreme defect (we also have to normalize for there being only one defect in our sample size). This defines $\sigma$.

\begin{equation}
\int_{\phi}^\infty \frac{x}{\sigma\sqrt{2\pi}}e^{-\frac{1}{2}\left(\frac{x}{\sigma}\right)^2} \mathrm{d}x=\frac{\sigma}{\sqrt{2\pi}}e^{-\frac{1}{2}\left(\frac{\phi}{\sigma}\right)^2}=\frac{1-\alpha^2}{2N}
\label{sigmadef2}
\end{equation}


To obtain an analytical expression, instead of solving equations (\ref{sigmadef1}) and (\ref{sigmadef2}), we can approximate. First, estimate that $\phi\approx\alpha$:

\begin{equation}
\int_{\alpha}^\infty \frac{1}{\sigma\sqrt{2\pi}}e^{-\frac{1}{2}\left(\frac{x}{\sigma}\right)^2} \mathrm{d}x=\frac{1}{2N}
\label{sigmasimp1}
\end{equation}

\begin{figure}[htbp]
\begin{center}
\includegraphics[width=0.9\linewidth,angle=0]{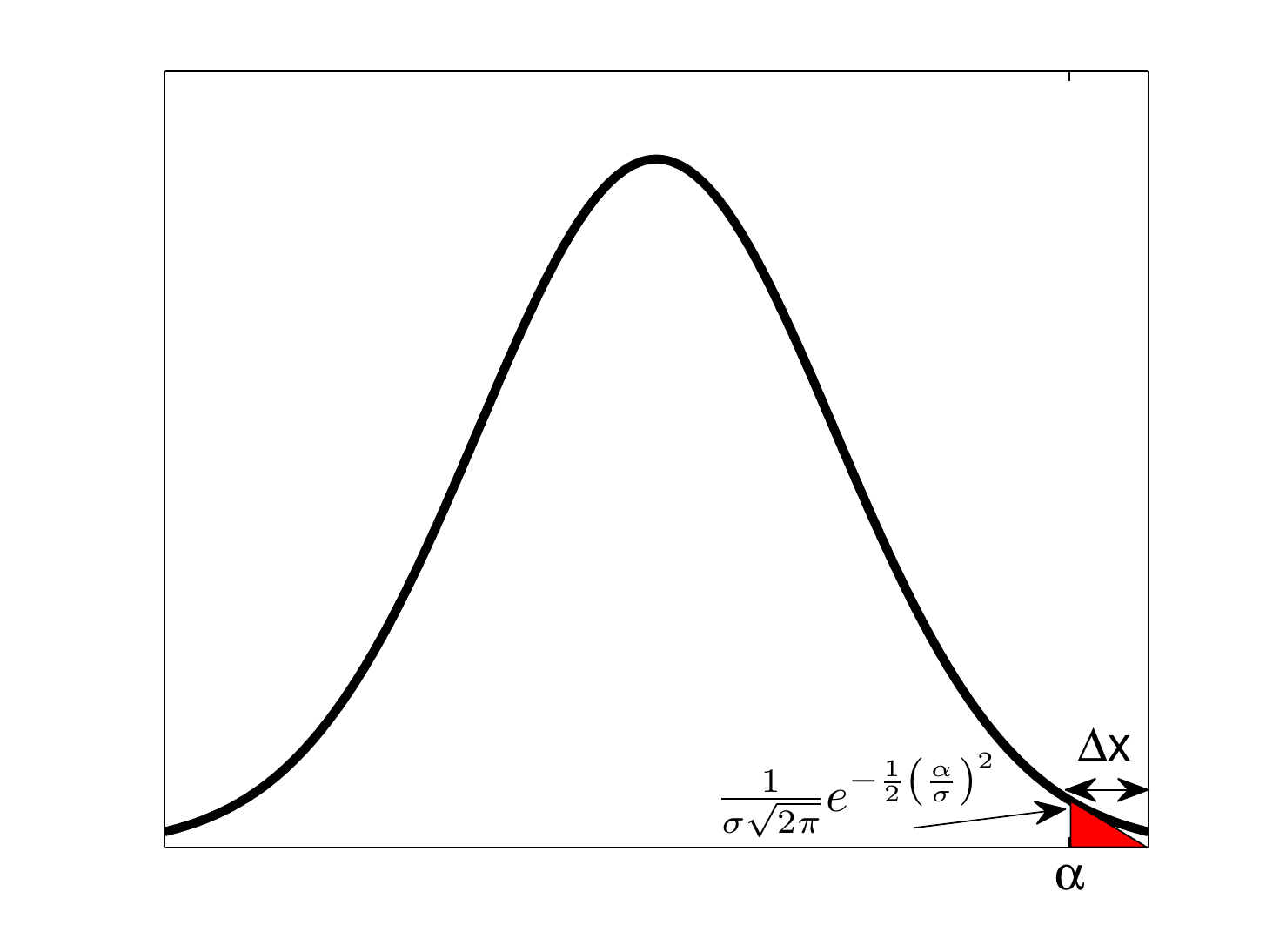}
\end{center}
\caption{Approximations to find an analytical expression for $\sigma$.}
\label{fig:meandering}
\end{figure}

We can also use a linear approximation for the Gaussian, so that the integral can be evaluated by calculating the area of the triangle shown in Fig.~(\ref{fig:meandering}). The equation of the line evaluated at $y=0$ determines the length $\Delta x$ in the figure:

\begin{equation}
0=\frac{-\sigma \alpha}{\sqrt{2\pi}}e^{-\frac{1}{2}\left(\frac{\alpha}{\sigma}\right)^2}\left(\Delta x\right)+\frac{1}{\sigma\sqrt{2\pi}}e^{-\frac{1}{2}\left(\frac{\alpha}{\sigma}\right)^2}
\label{sigmasimp2}
\end{equation}

This gives $\Delta x=\sigma^2/\alpha$. Evaluating the integral,

\begin{equation}
\int_{\alpha}^\infty \frac{1}{\sigma\sqrt{2\pi}}e^{-\frac{1}{2}\left(\frac{x}{\sigma}\right)^2} \mathrm{d}x=\frac{\sigma}{2\alpha\sqrt{2\pi}}e^{-\frac{1}{2}\left(\frac{\alpha}{\sigma}\right)^2}=\frac{1}{2N}
\label{sigmasimp3}
\end{equation}

Rearranging,

\begin{equation}
\frac{\alpha^2}{2\sigma^2}=\ln{\frac{\sigma N}{\alpha\sqrt{2\pi}}}
\label{sigmasimp4}
\end{equation}

N is very large, so $\ln{\frac{\sigma N}{\alpha\sqrt{2\pi}}}\approx\ln{N}$. Solving,

\begin{equation}
\sigma=\frac{\alpha}{\sqrt{2\ln{N}}}
\label{sigmasimp5}
\end{equation}

Now let us consider the behavior of the cavity at a field just above the expected vortex penetration field of the worst defect. Let $\hat{x}$ point into the film and $\hat{z}$ be aligned with the magnetic field. In this case, pairs of pancake vortices will form at the defect, move apart in $y$ due to the force exerted by the increasing magnetic field, and then move in the opposite direction in $y$ as the magnetic field direction reverses. In the mean time, they will have meandered some distance in $x$. If the meandering distance is very small compared to the impact parameter (approximately 50 nm), the pancake vortex pairs are most likely to meet again and annihilate. However, if the meandering distance is significant compared to the impact parameter, vortex pairs are likely to get lost, accumulating over the film.

For this calculation, we will not do a full simulation. Rather, we will consider a path along $y$ of a vortex pancake moving from the defect at $y=0$ to the extremum $y=y_{max}$, without any movement in $x$. We will, however, integrate the force in $x$ that the pancake would experience along its path, and calculate what the expected meandering distance would be from this.

The vortex will have two regions on either side of it in $x$ at a given time, one with $H_{c,1}$, and one with $H_{c,2}$. The centers of these two regions are separated by a distance $a$. The $x$-component of the force experienced by the vortex can be approximated from the gradient in condensation energy resulting from the difference in the local $H_c$. Magnetic energy density is given by $H^2/2\mu_0$. We convert this to energy by multiplying by the volume of a region, $a^2d$.

\begin{equation}
F_x=\frac{H_{c,1}^2-H_{c,2}^2}{2\mu_0a}a^2d
\label{forcedef}
\end{equation}

The motion of the pancake is described by Bardeen-Stephen drag:

\begin{equation}
\eta v_x=F_x/d
\label{BS}
\end{equation}

Here $v_x$ is the x-component of the velocity of the pancake and $\eta$ is the Bardeen-Stephen drag coefficient, defined by $\eta=H_{c2}\phi_0/\rho_n$, where $H_{c2}$ is the upper critical field of the film, $\phi_0$ is the flux quantum, and $\rho_n$ is the normal resistivity of the film.

The meandering distance $\Delta x$ within a region is given by $\Delta x=v\Delta t$, where $\Delta t$ is the time spent in that region. For simplicity, we approximate that the vortex moves with uniform speed in $y$, spending equal time in each of the regions that it travels through, such that $\Delta t=\frac{1}{2f}\frac{a}{y_{max}}$.

Using Eq. Eq.~(\ref{Bdist}) and Eq.~(\ref{forcedef}), we can describe a distribution $\mathcal{F}_x$ of forces that the pancake would experience. We observe that taking the difference between the absolute values from Eq.~(\ref{Bdist}) produces a normal distribution:

\begin{equation}
\mathcal{F}_x=\frac{a d H_{c,nominal}^2}{2\mu_0}(0\pm\sigma)
\label{Fdist}
\end{equation}

Between two regions that are adjacent in $x$, the pancake will experience a single value of the distribution of force values. After it travels a distance $a$, it will encounter a new pair of regions and therefore a new force value. We can use $\Delta x=v\Delta t$ and Eq.~(\ref{BS}) to find $\sqrt{\left\langle\Delta x^2\right\rangle}$, the RMS wandering distance traveled by the vortex over a distance $a$:

\begin{equation}
\sqrt{\left\langle\Delta x^2\right\rangle}=\sqrt{\left\langle F^2\right\rangle}\frac{\Delta t}{d\eta}
\label{x1dist}
\end{equation}

Multiply by the square root of the number of steps in a period $N_{steps}=\frac{2y_{max}}{a}$ to find the total RMS wandering distance $\sqrt{\left\langle\Delta X^2\right\rangle}$. Use Eq.~(\ref{Fdist}).

\begin{equation}
\sqrt{\left\langle\Delta X^2\right\rangle}=\frac{a d \sigma H_{c,nominal}^2}{2\mu_0}\frac{\Delta t}{d\eta}\sqrt{\frac{2y_{max}}{a}}
\label{xdist1}
\end{equation}

Since $\Delta t=\frac{a}{v_y}$, from Eq.~(\ref{BS}) and using $F_y=\frac{\phi_0 H}{\mu_0}$, we obtain:

\begin{equation}
\frac{\Delta t}{\eta}=\frac{a d}{F_y}=\frac{a d \mu_0}{\phi_0 \Delta H}
\label{dteta}
\end{equation}

Here $\Delta H$ is the difference in magnetic field across the film. If we are looking just above the penetration field, $\Delta H\approx\alpha \Hsh d/\lambda$. Eq.~(\ref{dteta}) can also be used to find $\frac{y_{max}}{a}$.

\begin{equation}
\frac{y_{max}}{a}=\frac{1}{2 f \Delta t} = \frac{\phi_0 \alpha \Hsh}{2 f \eta a \lambda \mu_0}
\label{ymaxa}
\end{equation}

Use the definition of $\eta$.

\begin{equation}
\frac{y_{max}}{a}=\frac{\rho_n \alpha \Hsh}{2 f H_{c2} a \lambda \mu_0}
\label{ymaxa2}
\end{equation}

Now substitute.

\begin{equation}
\sqrt{\left\langle\Delta X^2\right\rangle}=\frac{1}{2\sqrt{\ln{N}}}\frac{a^2 \lambda H_{c,nominal}^2}{\phi_0 \Hsh}\sqrt{\frac{\rho_n \alpha H_sh}{f H_{c2} a \lambda \mu_0}}
\label{xdist2}
\end{equation}

We then use $H_{c,nominal}=\frac{\phi_0}{2\sqrt{2}\pi\lambda\xi}$.

\begin{equation}
\sqrt{\left\langle\Delta X^2\right\rangle}=\frac{\xi}{4\sqrt{2}\pi\sqrt{\ln{N}}}\frac{H_{c,nominal}}{\Hsh}\sqrt{\frac{\rho_n \alpha \Hsh}{f H_{c2} a \lambda \mu_0}}
\label{xdist3}
\end{equation}

We can set our region size $a=\xi\approx 4$ nm. $\Hsh/H_{c,nominal}$ for Nb$_3$Sn is approximately 0.75. $N$ is approximately $10^{15}$, so $\frac{1}{4\sqrt{2}\pi\sqrt{\ln{N}}}$ is approximately $10^{-2}$. We also use $H_{c2}$=30 T, $\rho_n$=10 n$\Omega$m, and $H_{c,nominal}=530$ mT. For $\alpha=0.8$, these factors give $\sqrt{\left\langle\Delta X^2\right\rangle}\approx 2.1$ nm. Setting this as a standard deviation for vortex meandering and using $x_{imp}\sim$22 nm from above gives a result of 11 standard deviations, again making it extremely unlikely for vortices to escape due to disorder meandering.



Varying $a$ and $\alpha$ over a reasonable range also gives a small value for the meandering distance, as seen in Fig.~(\ref{fig:var}).

\begin{figure}[htbp]
(a) \par\smallskip
\centering
\includegraphics[width=0.9\linewidth]{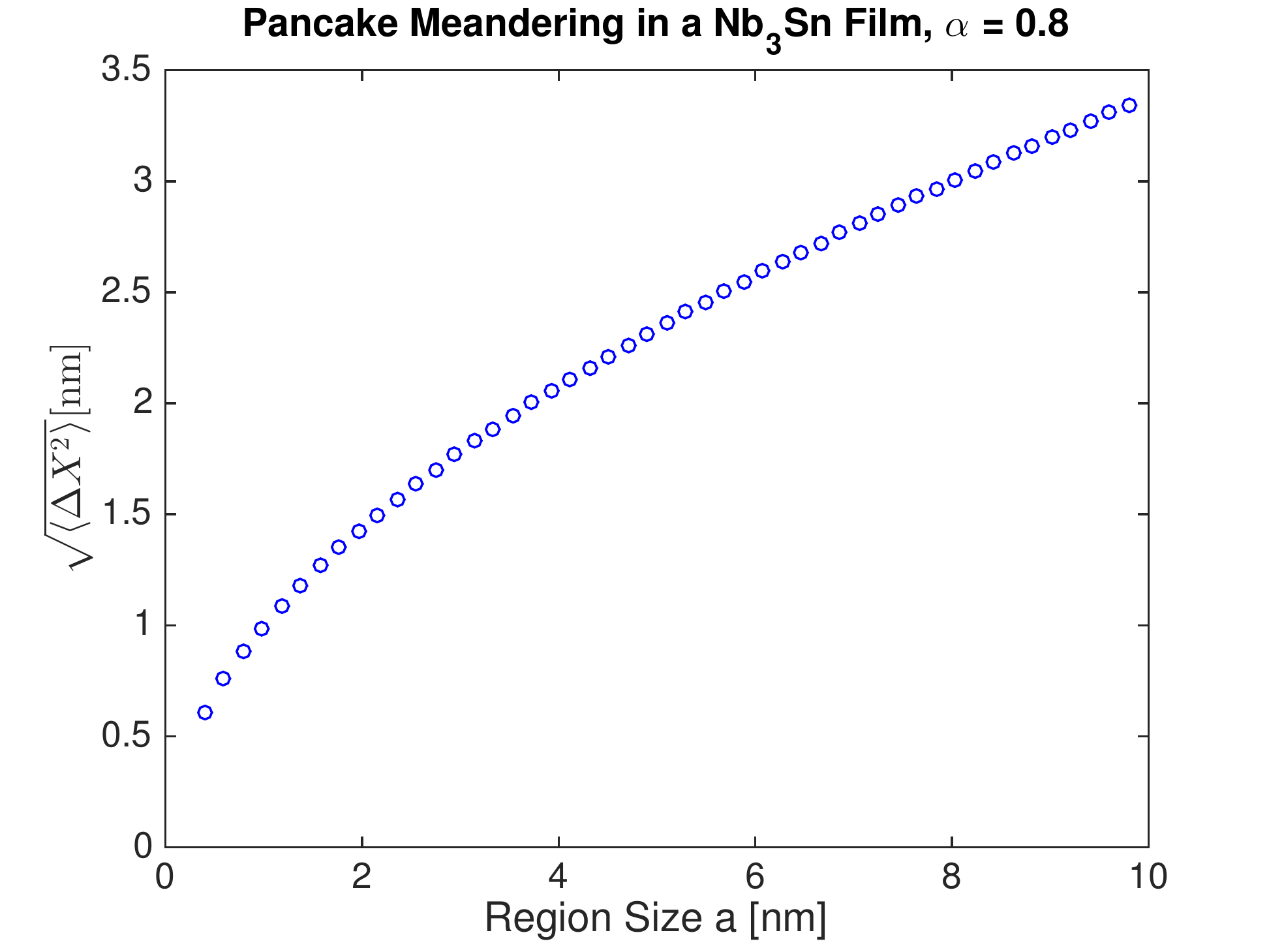}
\\
(b) \par\smallskip
\centering
\includegraphics[width=0.9\linewidth]{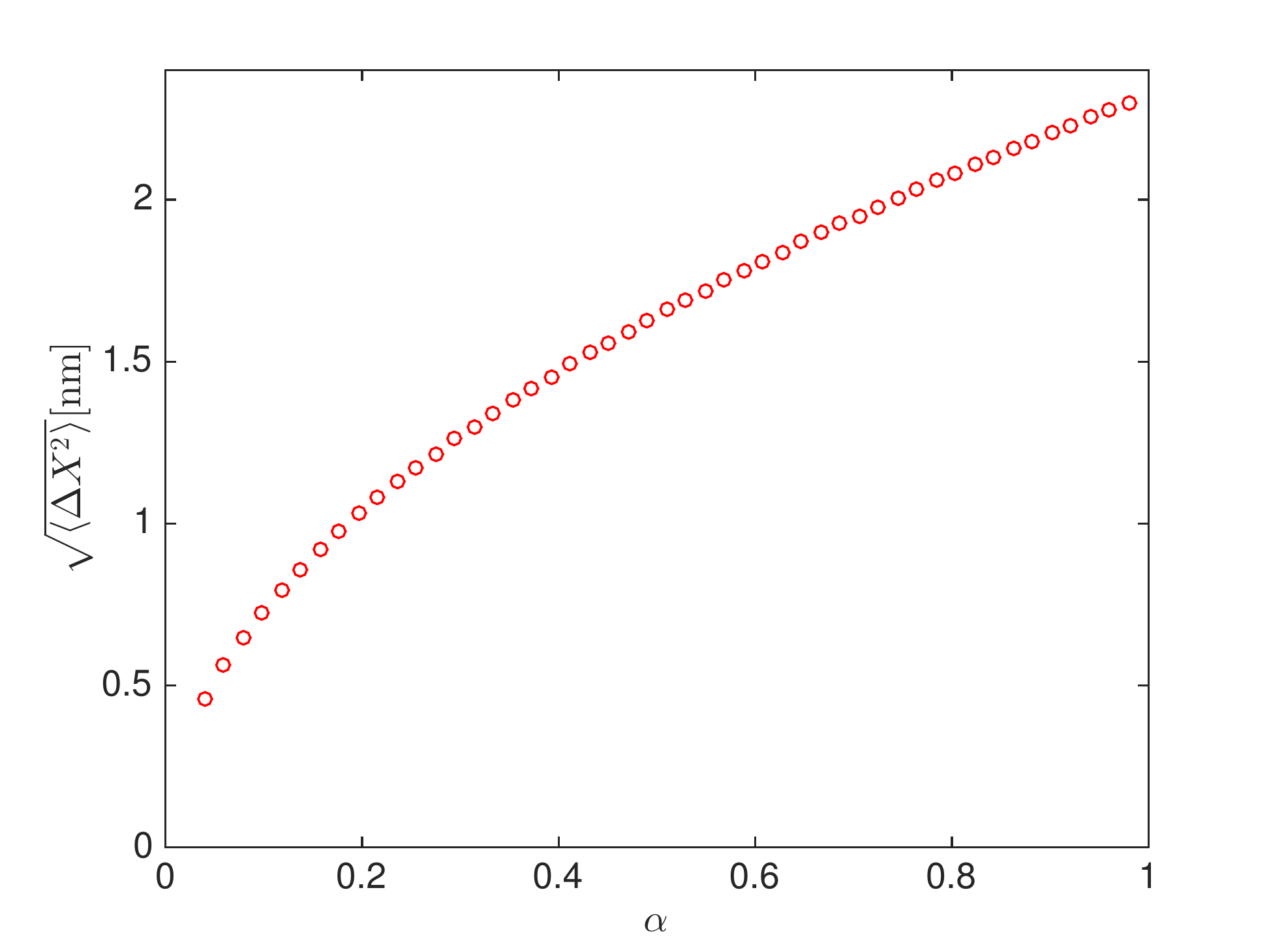}
\caption{The effect of varying $a$ for fixed $\alpha=0.8$ (a) and varying $\alpha$ for fixed $a=4$ nm (b) on the mean meandering distance of a pancake vortex due to disorder.} 
\label{fig:var}
\end{figure}

In addition to the impact parameter and thermal meandering distance, Fig.~(\ref{fig:wandercalc}) plots the meandering distance due to disorder using the analytical formulation above. Also plotted in the figure is a simulation of vortex pair creation and annihilation in the presence of disorder. In addition to forces from the applied magnetic field and from the randomly distributed array of circular defects with radius $a$,%
  \footnote{In the simulation, the defects pin vortices with force that increases linearly from the defect edge where it is zero, to the center of the defect, where it is a maximum. The maximum is set such that moving the vortex from the center of the defect to the edge would require work equal to the condensation energy of the volume of the vortex core.}
the simulation considers the forces of one vortex on the other, and finds the maximum lateral separation between the pair during a cycle.

\subsection{Interacting vortices, interacting defects}
\label{subsec:interacting}

We have presented analyses of thermal meandering and meandering of a
vortex-antivortex pair due to disorder, and so far, nothing has caused a
large buildup of vortices in the film. We have also performed a
preliminary analysis of the nucleation of many vortices at a defect over
the course of an RF cycle, all of which interact with each other. While
an attractive force exists between vortices and antivortices, vortices
exert a repulsive force on other vortices, and antivortices exert a
repulsive force on other antivortices. These repulsive forces between
similar vortex types appear to result in substantially larger lateral
movement, which could lead to higher rates of non-annihilation. An
example video is included as supplemental material~\cite{url:pancakevideo};
three frames of which are shown in Fig.~(\ref{fig:video}).
Here the film thickness is 30 nm, parameters are appropriate for Nb$_3$Sn
material, and the disorder is modeled as 60 pinning sites randomly spread over
0.4 square microns that exert a force of $F_{\mathrm{max}} \, k_i \, \rho$, where
$F_{\mathrm{max}}=B_c^2 \, \mu_0 \, \xi^2 \, d \, r_{\mathrm{pin}} /\, 2$,
$k_i$ is a value between 0 and 1
that is randomly chosen for each pinning site, and
$\rho=r_{\mathrm{sep}}/r_{\mathrm{pin}}$ for $r_{\mathrm{sep}}<r_{\mathrm{pin}}$
and 0 for $r_{\mathrm{sep}}>r_{\mathrm{pin}}$, where $r_{\mathrm{sep}}$
is the distance between the vortex and the center of the pinning site and
$r_{\mathrm{pin}}$ is the pinning site radius, (here chosen to be 2 nm).
Note that the maximum
horizontal meandering due to disorder and interactions is roughly eight
nanometers, only a factor of three less than the impact parameter
$x_{imp}$, suggesting that interactions may be much more dangerous
for multiple defects over many cycles. However, further work on interactions 
is clearly needed; it is possible that realistic parameters for the 
disorder strength could decrease the meandering, and it is possible that
the distribution of maximum meandering distances near a defect
over multiple cycles could be sub-Gaussian. On the other hand, it is
likely that two defects that
both nucleate pancake vortices and are close enough together that the
vortices interact could be dangerous even in the absence of disorder.

\begin{figure}[htbp]
\begin{center}
(a)\includegraphics[width=0.7\linewidth,angle=0]{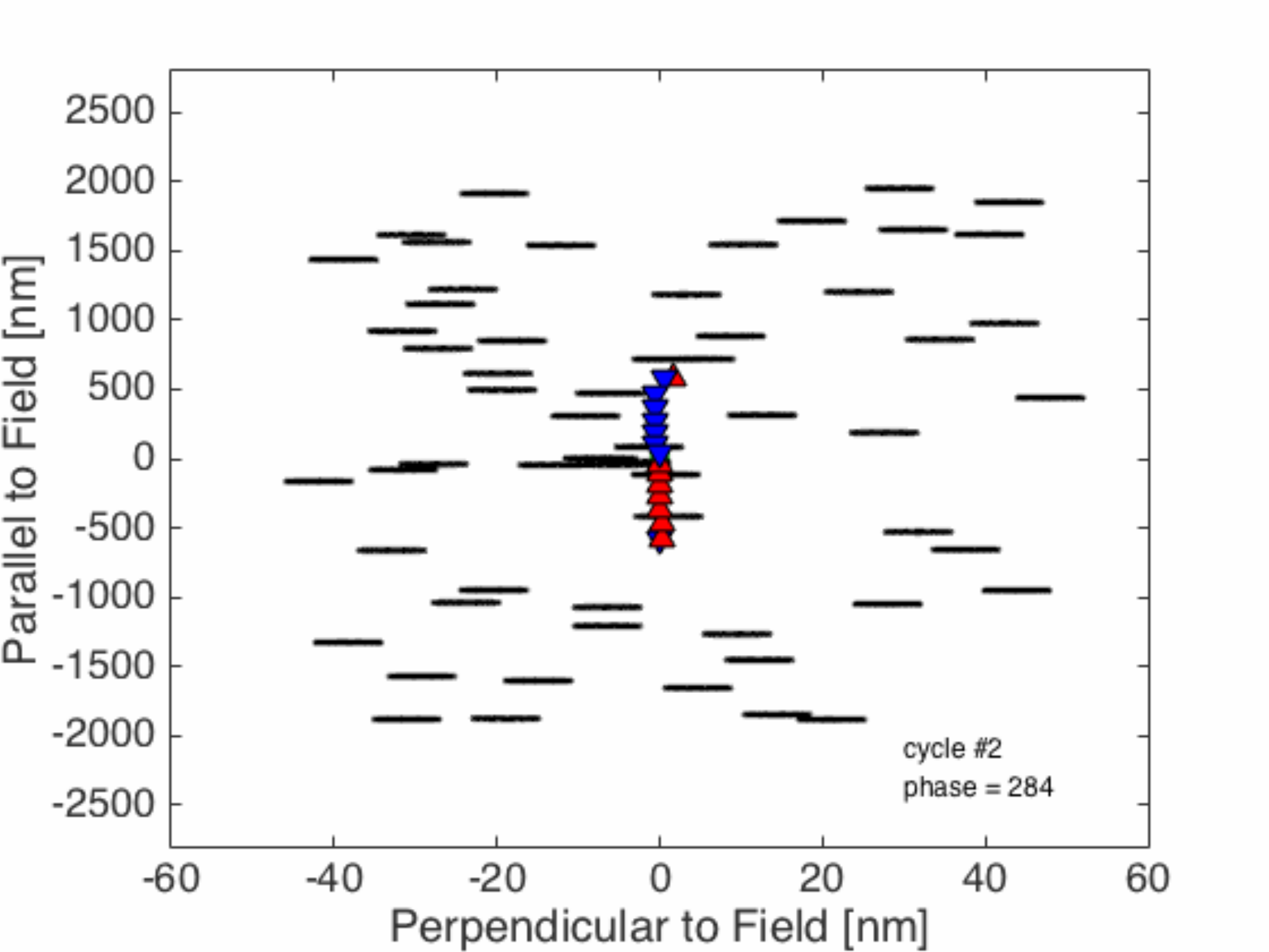}
(b)\includegraphics[width=0.7\linewidth,angle=0]{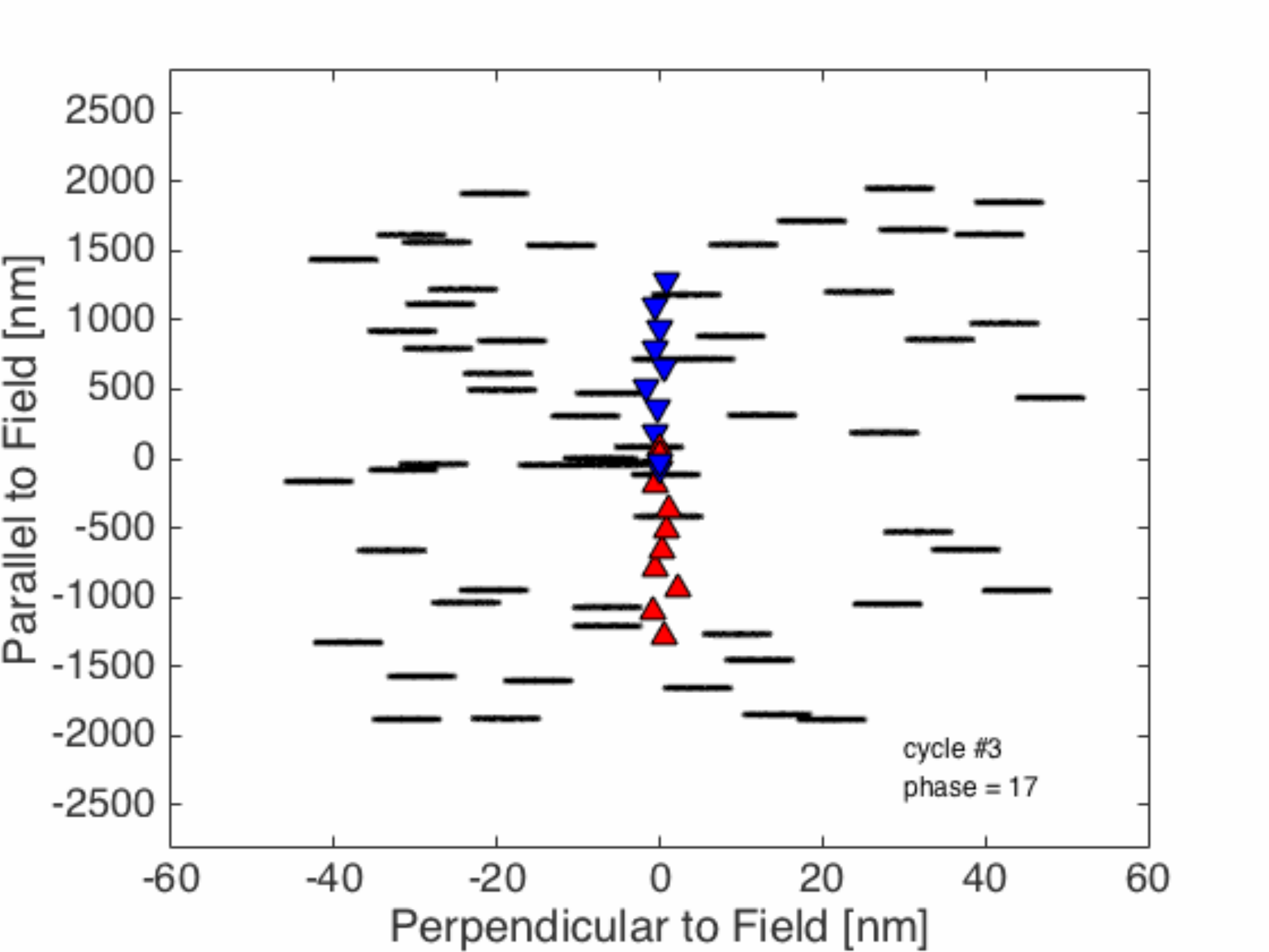}
(c)\includegraphics[width=0.7\linewidth,angle=0]{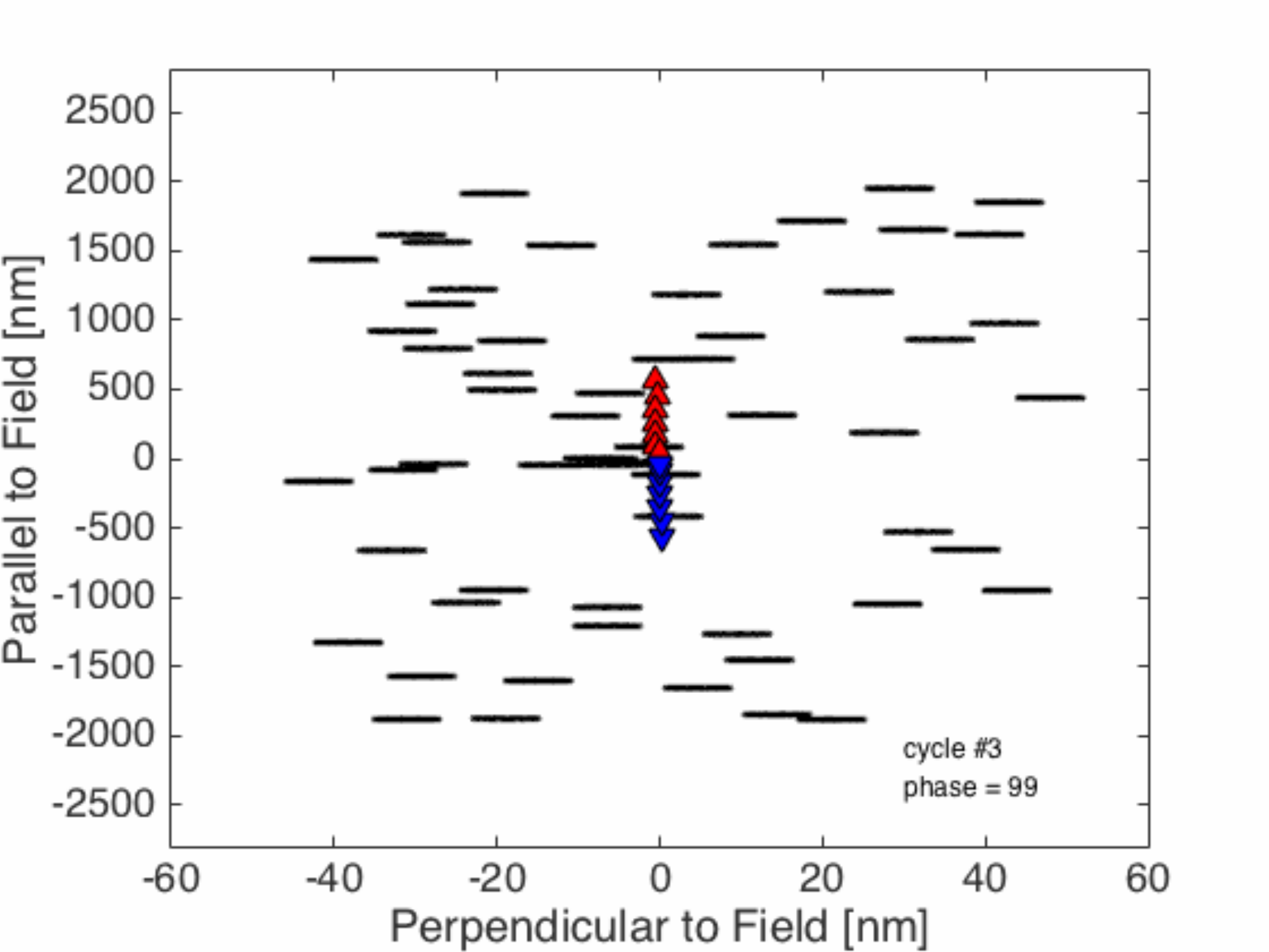}
\end{center}        
\caption{Frames from a preliminary simulation of pancake vortices (red) and antivortices (blue) being generated at a defect (green) in a thin film in an AC field.
The repulsion between similar vortices causes a lateral spread.
Note the horizontal and vertical scales differ by over a factor of 100.
The the horizontal lines are actually circular representations of the disorder 
potential; the vortices drift sideways much less than they traverse vertically.
(a)~Near zero field, (b)~Near a local maximum field, and (c)~Near a
return to zero field, showing the cycle-to-cycle variation.
}
\label{fig:video}
\end{figure}

%
%

%% file: Sec/Conclusions.tex
\section{Conclusions}
\label{sec:conclusions}

This article attempts to make a case for thoughtful, broad efforts to identify and estimate
fundamental physical limits of materials in parallel with experimental efforts. Our 
estimates for the superheating field of pure materials~\cite{catelani08,transtrum11}
had a significant impact in the SRF community, we understand, not because it promised 
that Nb$_3$Sn could be run at twice the field of Nb. Rather, we pointed out that the 
`vortex line nucleation' model was incorrect
(footnote~\ref{foot:Yogi} on page~\pageref{foot:Yogi})). This model,
created by and initially trusted in the SRF community, had provided discouraging estimates for high $\kappa$
materials, misleading the field for some years.
We also note that our controlled, theoretically grounded calculations
allow one to explore questions like materials anisotropy that cannot be gleaned reliably
from phenomenological models (or, indeed, from our own qualitative arguments,
Section~(\ref{subsec:anisotropy})).

We understand that many in the SRF community are skeptical of the use of bulk (or thick films)
of new materials with higher $\kappa$,
even though the theoretical estimates suggest significantly improved $\Hsh$
as well as lower cooling requirements. We too were concerned until recently 
that the smaller coherence lengths might make the metastable state more susceptible to 
vortex penetration. But we believe that our calculation of the effects of disorder within
a particular model clearly indicates that the reliability of the new materials increases so
rapidly away from $\Hsh$ that the effects of lower coherence length should not be a
major concern. One must always make choices of where to focus resources (laminates versus
bulk materials, coated copper versus niobium~\cite{russo09}; an interesting review has been recently 
published in this special issue~\cite{feliciano16}) -- but informed choices
may involve consultation with experts outside the SRF field.

We are also guardedly pessimistic about the utility of thin laminates in increasing
performance. (a)~We are concerned that experimentalists continue to be inspired
by the very high parallel fields sustainable by isolated thin films~\cite[Section 9]{feliciano16}
(footnote~\ref{foot:IsolatedLayer} on page~\pageref{foot:IsolatedLayer}), even though these
high fields are irrelevant in GHz applications~\cite{Hc1WrongRef} and likely also
practically inaccessible in any AC application. (b)~Dangerous vortices in thin films
will not typically reside parallel and inside the films,
but penetrate in and out of the film via pancake vortices, whose motion dissipates heat.
The maximum fields reachable without flux penetration for ideal thin films are rather 
similar to those of the bulk material~\cite{posen15}. The flux penetration needed to 
reach higher fields produces heating per cycle comparable to that for copper 
cavities~\cite{Hc1WrongRef}. (c)~We explore extensively the suggestion~\cite{Gurevich2006}
that the insulating layers in laminates may act to trap flux from defects, keeping flux from
entering the bulk. Here the key issue is whether the nucleated pancake vortices escape from the
vicinity of their parent flaw, or annihilate with their partners at the end of each cycle.
Modeling these as sources for pancake vortices, we agree that neither thermal diffusion
nor disorder are dangerous, but that interactions between vortices, and between vortices generated
at separate defects, might allow for escape and unacceptable heating -- warranting experimental
caution and further theoretical study.

Finally, a word to our colleagues outside the accelerator community. This paper is a collaboration
of SRF experts from the accelerator community (Posen, Liepe) and condensed matter theorists, and 
we draw heavily on conversations with both theorists and experimentalists inside the community
(Hasan Padamsee, Alex Gurevich, Nicholas Valles, Takayuki Kubo, Kenji Saito).
These domain-specific experts have enormous experience in the challenges and issues relevant for
the field; we were told that superheating fields, higher $\kappa$ materials, anisotropy, 
disorder, and laminates were the key questions, and have been guided into studying these in the
correct limits and focusing on the right issues. We can testify that this teamwork has led to
both excellent condensed-matter physics and efficient, targeted research to improve SRF performance.